\documentclass{emulateapj}
\usepackage{graphicx}

\makeatletter
\newcommand*{\rom}[1]{\expandafter\@slowromancap\romannumeral #1@}
\newcommand{\pasa}{PASA}
\usepackage[backref,breaklinks,colorlinks,citecolor=blue]{hyperref}
\newcommand{\OIII}{\hbox{[{\rm O}\kern 0.1em{\sc iii}]}}

 \newcounter{counter}       
 
 \newcommand{\lowerromannumeral}[1]{\setcounter{counter}{#1}\roman{counter}}

\makeatother
\shorttitle{Kinematic Downsizing at \lowercase{$z$}$\,\sim\,$2}
\shortauthors{Simons et al.}

\usepackage{verbatim}

\begin{document}

\title{Kinematic Downsizing at {\lowercase{$z$}}$\sim2$}

\author{Raymond C. Simons\altaffilmark{1}, Susan A. Kassin\altaffilmark{2}, Jonathan R. Trump\altaffilmark{3$\dagger$}, Benjamin J. Weiner\altaffilmark{4}, Timothy M. Heckman\altaffilmark{1}, Guillermo Barro\altaffilmark{5}, David C. Koo\altaffilmark{6}, Yicheng Guo\altaffilmark{6},   Camilla Pacifici\altaffilmark{7$\wedge$},  Anton Koekemoer\altaffilmark{2},  and Andrew W. Stephens\altaffilmark{8}}

\begin{abstract}
We present results from a survey of the internal kinematics of 49 star-forming galaxies at $z\sim2$ in the CANDELS fields with the Keck/MOSFIRE spectrograph (SIGMA, Survey in the near-Infrared of Galaxies with Multiple position Angles). Kinematics (rotation velocity $V_{rot}$ and integrated gas velocity dispersion $\sigma_g$) are measured from nebular emission lines which trace the hot ionized gas surrounding star-forming regions. We find that by $z\sim2$, massive star-forming galaxies ($\log\,M_*/M_{\odot}\gtrsim10.2$) have assembled primitive disks: their kinematics are dominated by rotation, they are consistent with a marginally stable disk model, and they form a Tully-Fisher relation. These massive galaxies have values of $V_{rot}/\sigma_g$ which are factors of 2--5 lower than local well-ordered galaxies at similar masses. Such results are consistent with findings by other studies. We find that low mass galaxies ($\log\,M_*/M_{\odot}\lesssim10.2$) at this epoch are still in the early stages of disk assembly: their kinematics are often supported by gas velocity dispersion and they fall from the Tully-Fisher relation to significantly low values of $V_{rot}$. This ``kinematic downsizing" implies that the process(es) responsible for disrupting disks at $z\sim2$ have a stronger effect and/or are more active in low mass systems. In conclusion, we find that the period of rapid stellar mass growth at $z\sim2$ is coincident with the nascent assembly of low mass disks and the assembly and settling of high mass disks.
\end{abstract}

\keywords{galaxies: evolution - galaxies: formation -galaxies: fundamental parameters - galaxies: kinematics and dynamics}

\affil{$^1$Johns Hopkins University, Baltimore, MD, 21218, USA; rsimons@jhu.edu\\
	$^2$Space Telescope Science Institute, 3700 San Martin Drive, Baltimore, MD, 21218, USA\\		
	$^3$Department of Astronomy and Astrophysics and Institute for Gravitation and the Cosmos, 525 Davey Lab, The Pennsylvania State University, University Park, PA 16802, USA\\
	$^4$Steward Observatory, 933 N. Cherry St, University of Arizona, Tucson, AZ 85721, USA\\	
	$^5$Department of Astronomy, University of California Berkeley, 501 Campbell Hall, Berkeley, CA 94720, USA\\
	$^6$UCO/Lick Observatory, Department of Astronomy and Astrophysics, University of California, Santa Cruz, CA 95064, USA\\
	$^7$Astrophysics Science Division, Goddard Space Flight Center, Code 665, Greenbelt, MD 20771, USA\\
	$^8$Gemini Observatory, Northern Operations Center, 670 N A'ohoku Place, Hilo, HI 96720\\
	$^\dagger$Hubble Fellow\\
	$^\wedge$NASA Postdoctoral Program Fellow}

\section{Introduction}

The cosmic star-formation rate peaks between $z\,=\,1.5$ and $z\,=\,2.5$ \citep{2014ARA&A..52..415M} and marks a critical period for galaxy assembly. In the classic model of galaxy formation, collisional baryons cool and dissipate into the center of dark matter halos while conserving primordial angular momentum, leading to the formation of a thin rotating disk galaxy \citep{1980MNRAS.193..189F, 1998MNRAS.295..319M}. Although this gas is expected to collapse to a disk in a few crossing times, external processes can disrupt or destroy disks on timescales shorter than the rotational period. In an emerging new picture, clumpy and smooth accretion from the cosmic web can efficiently carry external angular momentum inside the halo viral radius and directly onto the central galaxy \citep{2009ApJ...694..396B, 2011arXiv1106.0538K, 2013ApJ...769...74S, 2015MNRAS.449.2087D}, disrupting the velocity field \citep{2010ApJ...719.1230A, 2010ApJ...712..294E, 2012MNRAS.425..788G, 2015MNRAS.450.2327Z}. Additionally, ordered disks can be disturbed or destroyed through major mergers \citep{1996ApJ...471..115B, 2009ApJ...691.1168H, 2010ApJ...710..279C}, violent disk instabilities \citep{2009ApJ...703..785D, 2016MNRAS.456.2052I} and stellar feedback \citep{1999ApJ...513..142M, 2009ApJ...699.1660L,2015arXiv151101095W, 2015arXiv150900853A}.

These processes are prominent at $z\sim2$, coincident with the assembly of stellar mass in galaxies and may play a stronger role in the shallow potential wells that host low mass galaxies. In general, the formation and development of a galaxy depends strongly on its mass. Massive galaxies tend to assemble their stellar mass first (``downsizing", e.g., \citealt{1996AJ....112..839C, 2000ApJ...536L..77B, 2005ApJ...619L.135J}) and develop disks earlier (e.g., \citealt{2012ApJ...758..106K, 2014ApJ...792L...6V}). In contrast, low mass ($\log\,M_*/M_{\odot}\,\lesssim\,10$) galaxies tend to exhibit more disturbed and irregular morphologies at both low and high redshift (e.g., \citealt{2013MNRAS.433.1185M, 2015MNRAS.452..986S}), indicating that they lack mature structure. Recent evidence from imaging suggests that a significant fraction of low mass galaxies at $z\,>\,$1.5 appear to have elongated stellar distributions \citep{2014ApJ...792L...6V}. A similar phenomenon has been seen in hydrodynamic simulations of galaxy formation \citep{2015MNRAS.453..408C, 2016MNRAS.458.4477T} which indicate that oblate (stellar) disks form at late times ($z\lesssim2$), when the central potential becomes baryon-dominated. The star-forming gas in these simulated galaxies starts off in a highly perturbed and triaxial state, eventually collapsing to form thick oblate disks \citep{2015arXiv151208791M,2016MNRAS.458.4477T}. 

From $z\sim1.2$ to now, star-forming systems appear to have evolved through kinematic downsizing in a process known as ``disk settling" \citep{2012ApJ...758..106K}. In this picture, galaxies hierarchically settle from turbulent systems with high dispersion to regularly rotating disks, with more massive galaxies forming disks earlier. Locally: the majority of high mass galaxies are ordered rotating disks, with kinematically disordered galaxies appearing only below a stellar mass of $\log$ $M_*/M_{\odot}$ $<$ 9.5 \citep{2015MNRAS.452..986S}. However, it is unclear whether the kinematic downsizing picture extends beyond $z\sim1.2$ and into the peak of cosmic star-formation.

With recent advancements in near infrared detectors and instrumental multiplexing capabilities, space and ground based instruments have provided a wealth of knowledge regarding the physical properties of galaxies during this period, the cosmic noon. Ground based kinematic data for high redshift ($z\,\gtrsim\,$1) galaxies is rapidly growing, as reviewed in \citealt{2013PASA...30...56G}, with several recent and ongoing surveys: e.g., SINS \citep{2009ApJ...706.1364F}, WiggleZ \citep{2011MNRAS.417.2601W}, MASSIV \citep{2012A&A...539A..91C}, MOSDEF \citep{2015ApJS..218...15K}, KMOS3D \citep{2015ApJ...799..209W} and KROSS \citep{2016MNRAS.457.1888S}.

Emerging from these surveys is the general picture that star-forming galaxies at $z\sim2$ are {\emph {unlike}} the thin disk galaxies that make up the massive end up the local Hubble sequence. High redshift galaxies tend to be thick and gas-rich with velocity dispersions that are factors of a few higher than local thin disks. The typical velocity dispersion tends to increase with increasing redshift (e.g., \citealt{2008A&A...484..173P, 2012ApJ...758..106K}) and may reflect a similar rise in gas fractions \citep{2015ApJ...799..209W}. Even still, rotational support appears to be relatively common in the more massive galaxies ($M_*/M_{\odot}\gtrsim10$) at $z\sim2$, with $\sim\,75\%$ of the population exhibiting rotational motions comparable to or greater than their velocity dispersion \citep{2015ApJ...799..209W}. 

Only very recently, kinematic data for low mass galaxies ($\log\,M_*/M_{\odot}\,\lesssim\,10$) at high redshift is becoming available in gravitationally lensed systems (e.g., \citealt{2010MNRAS.404.1247J, 2015MNRAS.450.1812L}) and in larger numbers in the ongoing surveys with multi-object single slits (MOSDEF; \citealt{2016ApJ...819...80P}) and integral field spectrographs (KMOS-3D; \citealt{2015ApJ...799..209W}). While the imaging data at this epoch indicates that massive galaxies tend to be disk-like and the low mass galaxies tend to be irregular, similar conclusions have not yet been drawn from kinematic measurements, and it is unclear whether kinematic downsizing is in effect at $z\sim2$.

In this paper we aim to take advantage of the multiplexing capabilities and the high sensitivity of Keck/MOSFIRE multi-object near-infrared spectrograph \citep{2012SPIE.8446E..0JM} as well as the high resolution and sensitivity of Hubble Space Telescope (HST)/WFC3 H-band imaging to explore the kinematic state of star-forming galaxies at $z\sim2$. This survey is named SIGMA, Survey in the near-Infrared of Galaxies with Multiple Angles. The sample is morphologically diverse and is selected along the star-formation main sequence in two redshift bins from $1.3\,<\,z\,<\,1.8$ and $2.0\,<\,z\,<\,2.5$. 

In \S 2 we discuss the observations and sample selection. In \S 3 we describe measurements of the galaxy physical properties from multi-wavelength imaging. In \S 4 we detail the measurements of the gas-phase kinematics of our sample, and discuss multi-PA observations. In \S 5 we present the $z\sim2$ Tully-Fisher relation. In \S6 we examine trends between the kinematics and physical properties, and in \S7 we present a comparison of our galaxies with a model for marginally stable gas-rich disks. In \S 8 we present our conclusions and in the appendix we present simulations to determine the effects of spatial resolution on the measurement of kinematics. We adopt a $\Lambda$CDM cosmology defined with (h, $\Omega_m$, $\Omega_{\Lambda}$) = (0.7, 0.3, 0.7).

\section{Observations and Sample Selection}

\begin{figure}

\includegraphics[angle=0,scale=.43]{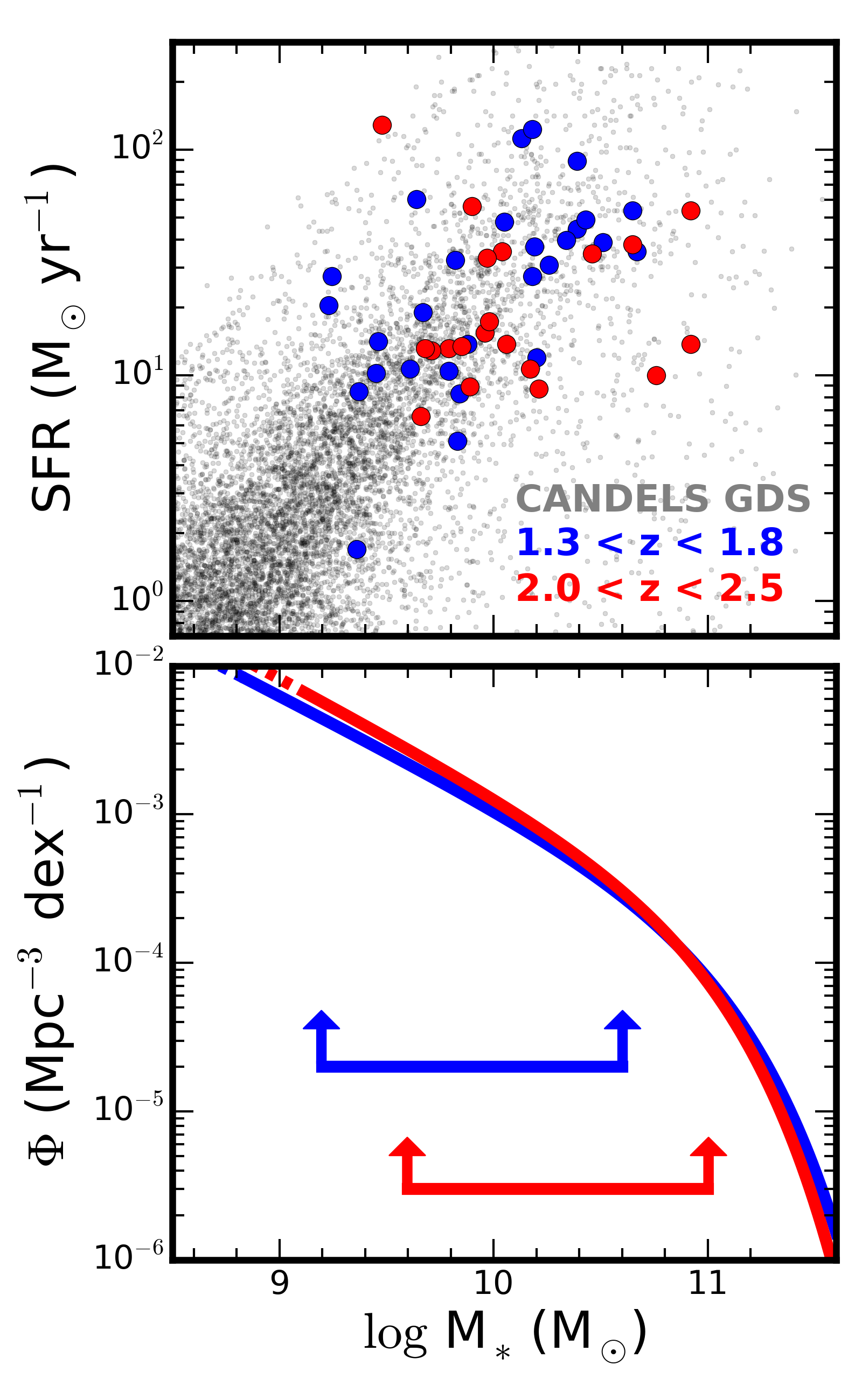}
\caption {Galaxies in SIGMA (large filled points) lie along the star-formation main sequence down to a stellar mass of $\log M_*/M_{\odot}\,=\,9.2$ at z $\sim$ 1.5 and $\log M_*/M_{\odot}\,=\,9.6$ at z $\sim$ 2.3. The grey dots represent galaxies drawn from the H-band selected GOODS-S CANDELS catalog in the same redshift range. A small amount of noise (0.05 dex) was added to the grey points to remove striations due to the discreet grid structure searched by the model fits to $M_*$ and star-formation rate (SFR). In the bottom panel, we compare the mass range of our sample, indicated by the red and blue arrows, with the blue galaxy stellar mass functions at the same redshifts, taken from \citet{2015MNRAS.447....2M}.}
\label{fig:M_sfr}
\end{figure}

The SIGMA sample is drawn from three of the fields (GOODS-S, GOODS-N and UDS) of the HST/WFC3 CANDELS survey \citep{2011ApJS..197...35G, 2011ApJS..197...36K}.  All of the spectra in GOODS-N were taken as a part of the TKRS-2 survey \citep{2015AJ....150..153W} and a subset of the spectra in GOODS-S and UDS have been presented in previous papers \citep{2013ApJ...763L...6T, 2014ApJ...795..145B}.  The SIGMA galaxy sample is shown in a star formation rate - stellar mass (SFR - $M_*$) diagram in Figure \ref{fig:M_sfr}.

This sample is a part of the well studied GOODS-N, GOODS-S, and UDS extragalactic fields, and so extensive multi-wavelength imaging is available for all of the galaxies in this sample. Galaxies in SIGMA are selected from the CANDELS WFC3/F160W catalogs \citep{2013ApJS..206...10G, 2013ApJS..207...24G} as follows. Selection priority was given to galaxies with spectroscopic redshifts from HST/WFC3 grism observations  \citep{2015AJ....149..178M}, taken as a part of the 3D-HST (\citealt{2012ApJS..200...13B, 2015arXiv151002106M}, PI Van Dokkum) and AGHAST (PI Weiner) surveys. The majority of the TKRS-2 galaxies were selected on previous spectroscopically confirmed redshifts. Photometric redshifts were derived from the broadband SEDs with EAzy \citep{2008ApJ...686.1503B, 2013ApJ...775...93D} and were used to select the remaining galaxies in our sample. Star-forming galaxies were selected down to a SFR of $\sim$3 M$_{\odot}$ yr$^{-1}$ at z $\sim$ 1.5 and $\sim$10 M$_{\odot}$ yr$^{-1}$ at z $\sim$ 2.2. There is no selection cut on morphology. As such, both regular and irregular galaxies are included in the sample. There is a priority weight on galaxy size, half-light V-band diameters of $>\,0.5\arcsec$, in order to select galaxies with emission extents that are larger than the typical seeing. 

The full SIGMA sample for which we have significant detections (S/N $>$ 3) in H$\alpha$ and/or \OIII$\lambda$5007 contains 97 galaxies. A portion of these were observed at multiple slit position angles (PA): 23 with 2 PAs and 11 with 3+ PAs. We select galaxies with at least one kinematic measurement coincident with the photometric major axis (within 45$^{\circ}$), removing 35 of the 97 galaxies. We make a further cut on emission line extent for each spectrum of each galaxy, as described in \S3.3 and the Appendix, such that the effective radius of the emission extent is at least 0.8 times the size of the seeing FWHM. This selection is shown in Figure \ref{fig:cont_size_em_size} and removes an additional 9 galaxies from the sample. An additional 4 galaxies are removed due to the presence of a strong overlapping skyline.

The final sample used in this paper contains 49 galaxies, 28 in the redshift range 1.3 $<$ z $<$ 1.8 and 21 at 2.0 $<$ z $<$ 2.5. The properties of the full sample are presented in Table \ref{tab:table1} of the Appendix.

\subsection{NIR Spectra}

The strong optical nebular emission lines (e.g., H$\alpha$, \OIII$\lambda$5007) from which we measure kinematics shift into the near infrared (NIR) at z $>$ 1.3. The multi-object spectrograph, MOSFIRE \citep{2012SPIE.8446E..0JM} on Keck-I, is one of the premiere instruments to study spectral features in the NIR. The spectra in this sample were taken in the H-band (1.46-1.81 $\mu$m) and K-band (1.93-2.45 $\mu$m). H$\alpha$ can be detected in the H-band from 1.3 $<$ z $<$ 1.8 and in the K-band from 2.1 $<$ $z$ $<$ 2.8 . Additionally, \OIII$\lambda$5007 falls in the H-band from 2.0 $<$ z $<$ 2.8. 

The MOSFIRE slit width was set to 0.7$\arcsec$ and the instrumental spectral resolution is R$\,\sim\,$3630 in both bands ($\sigma_{inst, HK}\,=\,35$ km s$^{-1}$). The masks were dithered along the slit between symmetric sky positions around the initial pointing to facilitate sky and instrument noise subtraction. The on-source exposure times were between 90 and 120 minutes with individual exposure times of 120 seconds.  The 2D spectra were sky-subtracted, wavelength-calibrated and rectified using the  MOSFIREDRP reduction pipeline\footnote{http://www2.keck.hawaii.edu/inst/mosfire/drp.html}. Spectroscopic redshifts were determined via Specpro \citep{2011PASP..123..638M}. 

Observations were taken in natural seeing conditions and the size of the seeing is estimated from unresolved calibration stars, where available, and marginally resolved compact ``blue nugget'' galaxies \citep{2014ApJ...795..145B} that were observed on the same mask as the galaxies in our sample. The NIR seeing ranged between 0.45 and 0.85$\arcsec$. Example cutouts of the 2D spectra around the H$\alpha$ emission line are shown in Figure \ref{fig:kin_example}.

\section{Quantities Measured from Imaging}

\subsection{Structural Parameters}

Throughout this paper we use two structural parameters measured from the HST/WFC3 images, namely the photometric position angle (PA, i.e., the direction of the photometric major axis) and the photometric axis ratio (b/a). For a well-behaved disk galaxy, the photometric PA will align with the kinematic PA (the direction of the largest velocity gradient) and the axis ratio can be used to correct the derived rotation velocity for the inclination of the galaxy to the line of sight. The photometric PA and axis ratio were measured with {\tt GALFIT} \citep{2010AJ....139.2097P} by \citet{2012ApJS..203...24V} on the H-band (F160W), J-band (F125W) and Y-band (F105W) for all of the H-Band selected CANDELS galaxies in our fields. We refer to these papers for measurement details. For this paper, we adopt the H-band measurements (rest I at $z\sim1.5$, rest V at $z\sim2.5$).

\subsection{Stellar Masses and Star-Formation Rates}

The available broadband measurements span from the UV to the NIR with ancillary Spitzer/MIPS 24/70$\micron$ photometry and Herschel far-IR photometry from the GOODS-Herschel \citep{2011A&A...533A.119E} and PACS Evolutionary Probe \citep{2013A&A...553A.132M} surveys.

As described in \citet{2014ApJ...795..145B}, integrated stellar masses ($M_*$) were derived from the broadband SED of each galaxy using the FAST fitting code \citep{2009ApJ...700..221K} assuming a \citet{2003PASP..115..763C} initial mass function, \citet{2003MNRAS.344.1000B} stellar population synthesis models and a \citet{2000ApJ...533..682C} extinction law. The errors on the stellar masses are $\sim\,$0.3 dex \citep{2015ApJ...808..101M}. Dust-corrected star-formation rates were derived following the SFR-ladder \citep{2011ApJ...738..106W}. For galaxies with Spitzer and Herschel far-IR detections, the SFR is calculated from both the obscured (IR) and unobscured (UV) components (following \citealt{1998ARA&A..36..189K}). For non-detections in the mid-far IR, the star-formation rates are derived from the UV and are extinction-corrected following the best fit attenuation from the SED modeling. In Figure \ref{fig:M_sfr} we show the star-formation rate vs stellar mass diagrams for our sample, the ``star-formation main sequence" (e.g. \citealt{2007ApJ...660L..43N}). Galaxies drawn from the full CANDELS catalog in GOODS-S are shown by black dots in the background. Our sample is generally representative of the SF main sequence down to a stellar mass $\log\,M_*/M_{\odot}\,=\,9.2$ in the 1.3 $<$ z $<$ 1.8 redshift bin and 9.6 in the 1.8 $<$ z $<$ 2.8 bin. We make a comparison with the blue galaxy stellar mass functions at 1.5 $<$ z $<$ 2.0 and 2.0 $<$ z $<$ 2.5 \citep{2015MNRAS.447....2M} and note that our sample spans close to the knee of the Schechter function, $M^*\,\sim\,10^{11}\,M_{\odot}$, in both redshift intervals.

 \begin{figure}
\begin{centering}
\includegraphics[angle=0,scale=.55]{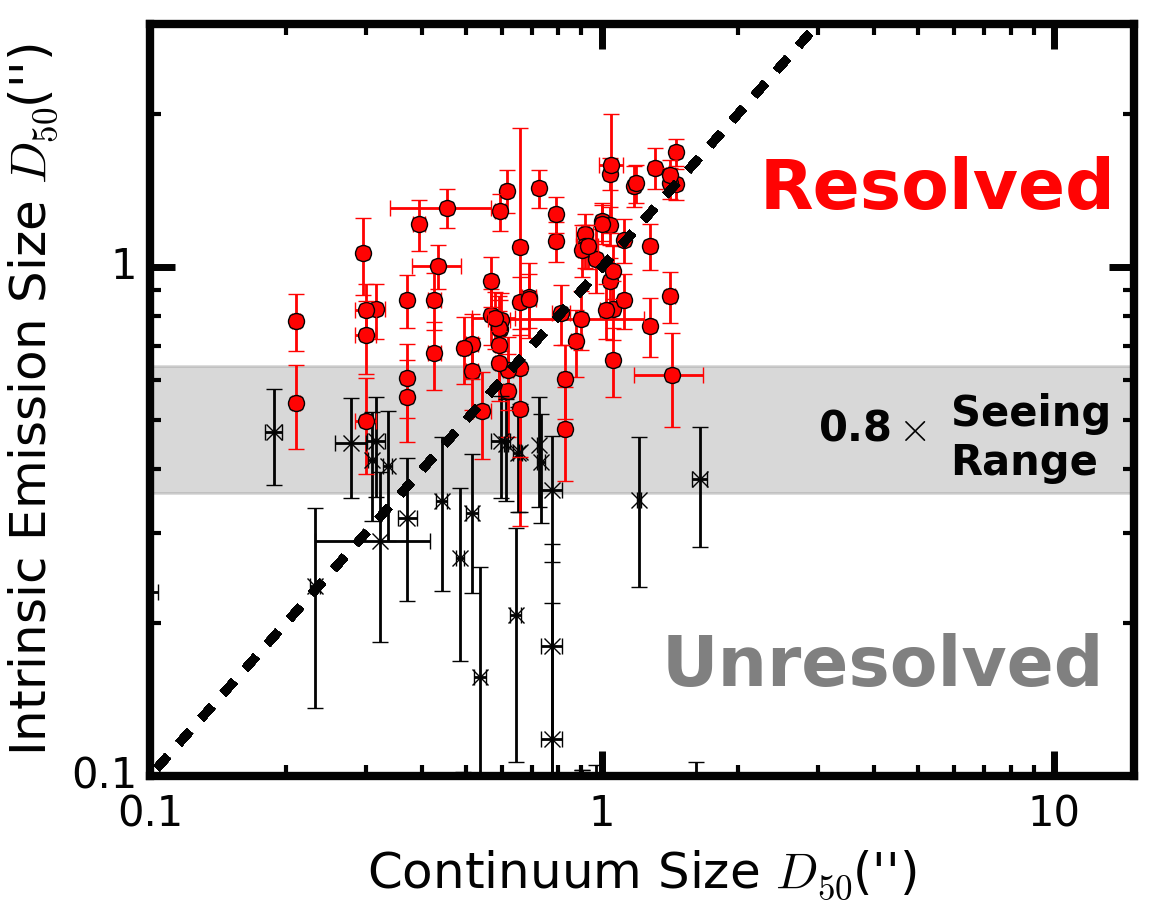}
\caption {Spatially resolved observations are identified from the intrinsic extents of their emission lines. We plot the intrinsic half-light diameter ($D_{50}$) of the emission versus the continuum for each slit on every galaxy used in this paper. Resolved observations are marked with red filled circles and the discarded unresolved observations are marked as black x's. The range of seeing conditions during the observations is shown as a grey shaded swath. Emission line sizes are seeing corrected from the MOSFIRE 1D emission profile. Continuum sizes are measured from Sersic fits to the H-band Hubble images. As shown in the Appendix, kinematic structure can be resolved down to a limiting resolution of D$_{50}$ = 0.8$\times$ Seeing$_{FWHM}$, where D$_{50}$ is the 1D seeing-corrected emission line size.  \vspace{0.1cm}}
\label{fig:cont_size_em_size}
\end{centering}
\end{figure}

 \subsection{Continuum and Emission Sizes}

One of the main challenges to measuring rotation with ground based data comes from the effects of beam smearing \citep{1987PhDT.......199B}. Seeing smooths out velocity gradients and artificially boosts the velocity dispersion in the central parts of rotating galaxies. This effect can generally be modeled, except in the cases where the emission extent is comparable to or smaller than the seeing. Galaxies with small emission extents can appear dispersion-dominated in seeing-limited data but reveal rotation at higher resolution \citep{2013ApJ...767..104N}. 

In order to examine the effects of beam-smearing on our data, we produce a suite of simulated MOSFIRE/SIGMA spectra in the Appendix. We conclude that, for the typical signal to noise of the spectra in our sample, we can reliably recover intrinsic kinematics ($V_{rot}$, $\sigma_g$) down to D$_{50}$ $\approx$ 0.8 $\times$ Seeing$_{FWHM}$, where D$_{50}$ is the 1D half-light diameter measured from the emission lines and Seeing$_{FWHM}$ is the full width at half maximum of the seeing.

We make size selections using the emission rather than the continuum for a couple of reasons. First, the 2D half-light radii of the emission typically extend beyond the continua \citep{2015arXiv150703999N}. Secondly, seeing tends to blur emission that is not coincident with the slit placement into the slit \citep{2006ApJ...653.1027W}. As such, an emission size determined from a projection of the slit onto the HST image may underestimate the extent of the emission measured with the slit.  By measuring sizes directly from the seeing-corrected emission lines, we bypass both of these uncertainties.

In Figure \ref{fig:cont_size_em_size} we show the half-light diameter sizes, in both emission and continuum, for the galaxies in our sample. For each slit, the emission line size is measured directly from the spectrally-collapsed emission line. Due to the convolution of the intrinsic profile with the seeing, the spatial profiles of the spectrally-collapsed emission lines are generally well described by a 1D Gaussian. The seeing is subtracted from the observed profile in quadrature ($\sigma_{em, intrinsic} = \sqrt{\sigma^2_{em, observed} - \sigma_{seeing}^2}$) and the half-light size is measured from the recovered intrinsic profile.

Continuum radii are measured from the HST/WFC3 H-Band image using the {\tt GALFIT} software package in \citet{2012ApJS..203...24V}. For the galaxies in our sample, the half-light radii of the intrinsic emission profiles tend to be slightly larger than the continuum sizes. This could be the case for two reasons. First, the half-light radius that is measured by {\tt GALFIT} is a 2D quantity measured on concentric ellipses while the emission size is being measured directly from the 1D emission lines, an along-the-slit measurement. Since the 1D measurement essentially integrates the light across the slit, these two sizes will not directly scale to one another in general. The difference may also be physical, in the sense that the emission profile can extend beyond that of the continuum (\citealt{2015arXiv150703999N}).

\begin{figure*}
\begin{centering}
\includegraphics[angle=0,scale=.68]{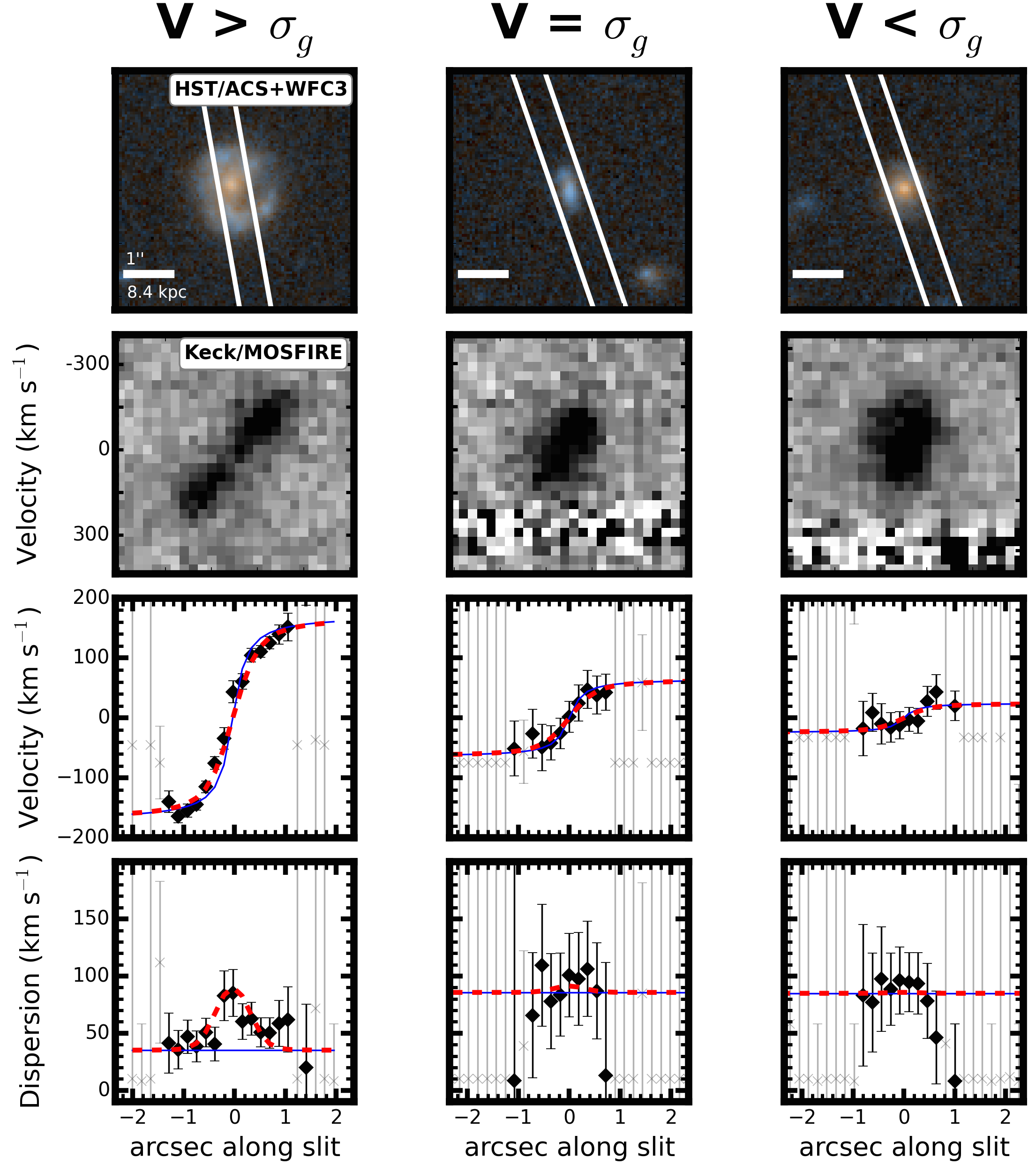}
\caption {Example single slit observations are presented for three SIGMA galaxies (ID: 16600, 16209, 14602). These galaxies span the kinematic types in our sample: a rotation dominated galaxy (left column), a galaxy with equal contributions of rotation and dispersion (middle column), and a dispersion dominated galaxy (right column). In the top two rows we show the I+H-band HST/ACS-WFC3 color images with the MOSFIRE slit placement and the 2-D spectra centered around the H$\alpha$ line. Strong NIR atmospheric lines are present in the middle and right columns. In the bottom two rows we show the kinematic model fits to the emission lines. The black filled diamonds represent the gaussian fits to the velocity and velocity dispersion in each row of each spectrum. The grey points are poor fits and are discarded. The best-fit models are shown as red solid lines and the intrinsic (pre-seeing blurred) models are shown as blue dashed lines. All of the rows are spatially aligned and each panel is 4.5$\arcsec$ on a side. Kinematic fits for all of the SIGMA galaxies are available in the Appendix.}
\label{fig:kin_example}
\end{centering}
\end{figure*}

\begin{figure*}
\begin{centering}
\includegraphics[angle=0,scale=.50]{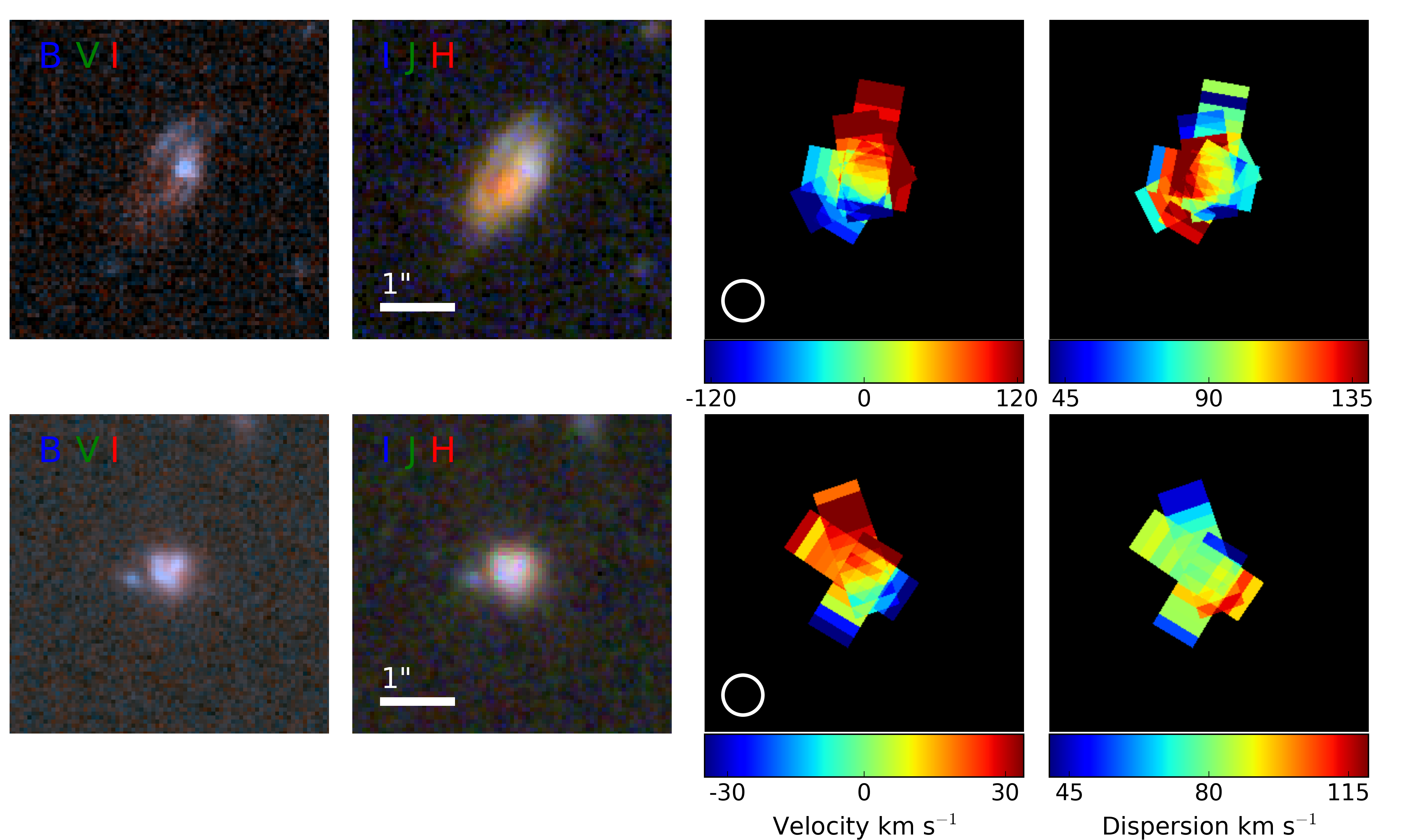}
\caption {Hubble images and reconstructions of 2D kinematic fields from multi-slit observations are shown for two SIGMA galaxies (ID: 26736, 21162). In the first two columns we show the the HST/ACS optical (rest UV) and the HST/WFC3 near-IR (rest optical) color images, respectively. In the third and fourth columns, we show the co-added velocity ($V\,\times\,\sin i$) and velocity dispersion maps, respectively. The top row shows a regularly rotating disk galaxy: the velocity field has a smooth rotation gradient with a centrally peaked velocity dispersion, both kinematic signatures of a rotating disk. The bottom row shows a dispersion dominated galaxy with complex morphology and kinematics: it exhibits velocity gradients in orthogonal slits. Each slit pixel is 0.18$''$ and the slits are 0.7$''$ wide. The typical seeing in the H-band is 0.6$''$, as shown by the white circles on the velocity maps.}
\label{fig:vel_map_combined}
\end{centering}
\end{figure*}

\section{Kinematics measured from emission lines}

Kinematics are measured from strong nebular emission lines (H$\alpha$, \OIII$\lambda$5007]) in the spectra. These lines generally trace the hot (T$\,\sim\,$10$^{4}$ K) ionized gas in HII regions surrounding ongoing star-formation. The kinematics are measured using the code {\tt ROTCURVE} \citep{2006ApJ...653.1027W}. The instrumental resolution of our spectra (R$\,\sim\,$3630) allow us to measure the velocity dispersion down to $\sim\,$20 km s$^{-1}$ and the rotation velocity uncorrected for inclination ($V_{rot} \times \sin i$) to $\sim\,$15 km s$^{-1}$. 
 
The kinematic modeling performed by {\tt ROTCURVE} has been described in detail in previous papers, so we will briefly outline it here and refer to \citet{2006ApJ...653.1027W} for further details. {\tt ROTCURVE} models the spatial profile of the emission line as a 1D Gaussian. In each row along the slit, we recover a velocity and velocity dispersion using an uncertainty-weighted least squares fit to a 1D Gaussian. Nearby, but not overlapping, skylines are manually masked to prevent noise peaks from driving the fit. The velocity structure of the line is modeled with two components: a rotation curve and a dispersion term. Due to our limited spatial resolution, the velocity dispersion is kept constant with radius in the model. The rotation velocity is modeled with an arctan function:

\begin{equation}
V_{x}= V\frac{2}{\pi}\arctan(x/r_v)
\end{equation}

\noindent
where r$_v$ is the knee radius for the turnover of the rotation curve and $V$ is the velocity of the flat part of the rotation curve. We note that the flat part of the rotation curve is not reached in a fraction of our galaxy sample (see Appendix) and in these cases $V$ is driven by the scale of the velocity difference reached by the data and the modeled turnover radius. Due to the seeing blur, r$_v$ is not well constrained and so we fix the turnover radius in the model at a fiducial value of 0.2$\arcsec$ (1.7 kpc at $z\sim2$) and adjust to 0.3$\arcsec$ or 0.4$\arcsec$ for galaxies that are not well fit initially. The turnover radius and the rotation velocity are moderately covariant with a $\pm\,$0.1\arcsec change in r$_v$ resulting in a $\pm\,$0.1 dex change in $V$. The model parameters of $V$ and $\sigma_g$ are explored on a grid of 5 km s$^{-1}$. The modeled emission line is then spatially  convolved with a fixed seeing kernel (0.6\arcsec\, FWHM), which is the typical seeing of our observations in the NIR. The best fit model is determined through a $\chi^2$ minimization with the data over this grid.

The velocity dispersion term that we measure ($\sigma_g$) is unlike a pressure supported stellar velocity dispersion. The hot gas which $\sigma_g$ is tracing can collisionally radiate and so a high dispersion system cannot remain in equilibrium after a crossing time. Due to the seeing blur, the velocity dispersion is a measure of the small scale velocity gradients below the seeing limit. The velocity dispersion measured from nebular emission lines will be a combination of: (\lowerromannumeral{1}.) thermal broadening of the gas ($\sim$10 km s$^{-1}$ for H at 10$^4$ K), (\lowerromannumeral{2}.) internal turbulence in HII regions ($\sim$ 20 km s$^{-1}$ for local HII regions; \citealt{1990ARA&A..28..525S}), (\lowerromannumeral{3}.) relative motions between HII regions.

The inclination is measured as $\cos^2i=((a/b)^2-\alpha^2)/(1-\alpha^2)$, where $\alpha$ is the correction for the thickness of the disk ($\alpha$ = 0.2) and b/a is the axis ratio of the galaxy, measured from the HST H-band image. The rotation velocity is then corrected as $V_{rot}\,=\,V/\sin i$. 

As indicated by \citet{2015ApJ...803...62H, 2016ApJ...816...99H}, distinguishing late-stage mergers and rotating disks is difficult in low resolution kinematics observations. At high redshift, where galaxy mergers are expected to be common \citep{2015MNRAS.449...49R},  disk classifications made from kinematic signatures alone may be biased high. In 2 galaxies in our sample (ID: 21897 and 4476) we identify an interacting neighbor in both the HST image ($<\,1\arcsec$) and the MOSFIRE spectrum ($<\,100\,$km s$^{-1}$). We mask the emission line of the neighboring source in both of these cases. We note that late-stage interactions, for which two galaxies are identified as a single source, may be included in this sample. A joint analysis of the HST imaging and the kinematic data is necessary to disentangle disks from late-stage mergers in our sample. For all subsequent analyses, we interpret the measured velocity gradient of each galaxy as rotation.

In Figure \ref{fig:kin_example}, we show example kinematic fits for three galaxies in our sample: a strongly rotating galaxy, a galaxy with comparable velocity and velocity dispersion, and a dispersion dominated galaxy. Even in the dispersion dominated case, we note the presence of a velocity gradient, although it is small compared to the observed dispersion. The rotationally supported system is a well-ordered disk galaxy, similar to disk galaxies today, with the presence of spiral structure and a bar in the HST image. In its spectrum, we find a strong rotation signature ($\sim$ 175 km s$^{-1}$) and a mild velocity dispersion ($\sim$ 35 km s$^{-1}$).

Beam smearing tends to blur together spatially separate regions of emission, leading to a flattening of nearby velocity gradients and an elevation in the local velocity dispersion. In the observed line, this effect leads to a classic peak in the velocity dispersion at the center of the galaxy, where the velocity gradient is the steepest. This can be seen in the example kinematic fits in Figure \ref{fig:kin_example}, particularly for the strongly rotating galaxy. {\tt ROTCURVE} models the seeing and accounts for this artificial beam smearing when constructing its model. 

The best-fit kinematic parameters of the full sample are presented in Table \ref{tab:table2} of the Appendix.

\subsection{Multiple Position Angle Observations}

Single-slit spectroscopy is a reliable method for recovering rotation velocity and velocity dispersion in observations where the position angle of the slit is aligned with the kinematic major axis of the galaxy. In ordered disk galaxies, the kinematic and photometric major axes are aligned and so the HST/WFC3 image can be used to determine the kinematic major axis. However, the morphologies of high redshift star-forming galaxies can be more severely  disturbed than local or low redshift samples (e.g., \citealt{2013MNRAS.433.1185M}), implying that the photometric and kinematic major axes might be misaligned. In a sample that combines galaxies at both $z\sim1$ and $z\sim2$, \citet{2015ApJ...799..209W} report that kinematic and photometric axes are at least moderately misaligned ($>\,15^{\circ}$) in $\sim$40$\%$ of galaxies and are largely misaligned ($>\,30^{\circ}$) in $\sim$20$\%$. This fraction declines to $<10\%$ in local galaxies \citep{2015A&A...582A..21B}.

For seeing-limited observations of high redshift galaxies, slit placement is much less stringent than for local galaxies or for higher resolution observations. To recover rotation, the alignment between the slit and the kinematic major axis should be within 45$^{\circ}$ (see Figure 13 of \citealt{2006ApJ...653.1027W}). 

A total of 26 galaxies in SIGMA have observations at multiple position angles ($PA_{slit}$, see Figure \ref{fig:vel_map_combined}). Due to the non-uniform coverage of our maps, we do not simultaneously model all of the observed slits on a galaxy. Instead we examine the kinematic measurements along each slit separately and adopt measurements for the slit with the highest rotational signature. We require that every galaxy in SIGMA have at least one slit that is aligned within 45$^{\circ}$ of the photometric major axis ($PA_{phot}$). This requirement ensures that rotation will be recovered in disk galaxies, where the photometric and kinematic axes are aligned. Of the 26 galaxies with kinematic measurements at more than one PA, 17 have at least one slit aligned and one slit misaligned with ${PA}_{phot}$. Of these systems, we find the largest velocity gradient in a misaligned slit in 23$\%$ of the cases (4 galaxies) and in an aligned slit in 77$\%$ of the cases (13 galaxies). This result is consistent with \citet{2015ApJ...799..209W}. The 4 galaxies in our sample for which the strongest velocity shear is found in a misaligned slit all display complex morphology in both the HST/ACS V-band and HST/WFC3 H-band images. 

For these 4 galaxies, we adopt the line of sight velocity shear ($\Delta V/2$) for our analysis and do not correct the velocity for inclination. These galaxies are marked in all of the figures. As shown in an example in Figure \ref{fig:vel_map_combined}, these systems tend to have complex velocity profiles. In some of these cases, velocity shear is found in the aligned slit as well, implying that the velocity gradient (either tracing global rotation or more complex motions) may lie between both slits.

\section{The Tully-Fisher Relation}

\subsection{Background}

The Tully-Fisher relation (TF; \citealt{1977A&A....54..661T}) is an empirical scaling relating the dynamics and the luminosity, or more recently, stellar mass of galaxies. In the local universe, the relation is relatively tight for massive disk galaxies (e.g., \citealt{ 2001ApJ...563..694V, 2001ApJ...550..212B, 2005ApJ...633..844P, 2006ApJ...643..804K, 2007ApJ...671..203C, 2008AJ....135.1738M, 2011MNRAS.417.2347R}) but begins to show small residuals to low mass for gas-rich galaxies \citep{2005ApJ...632..859M, 2016arXiv160202757B} and large residuals to low rotation velocity for irregular low mass galaxies ($\log\,M_{*}/M_{\odot}\,\lesssim\,9.5$)\citep{2015MNRAS.452..986S}. 

Numerous studies have extended the analysis of the TF relation to both intermediate, 0.1 $<$ z $<$ 1.3  (e.g., \citealt{2005ApJ...628..160C, 2006A&A...455..107F, 2006ApJ...653.1049W, 2007ApJ...660L..35K, 2008A&A...484..173P, 2011ApJ...741..115M, 2011MNRAS.416.1936T, 2011ApJ...741..115M, 2012ApJ...758..106K, 2012A&A...546A.118V, 2014ApJ...782..115M, 2015arXiv151200246C}), and high redshift, 1.3 $<$ z $<$ 3.0 (e.g., \citealt{2009ApJ...697..115C, 2012ApJ...753...74M, 2012MNRAS.426..935S, 2015MNRAS.450.1812L}).

\citet{2006A&A...455..107F} and \citet{2006ApJ...653.1027W} were the first to note that intermediate redshift galaxies with perturbed/complex kinematics and irregular morphologies tend to exhibit large scatter from the TF relation, with subsequent studies confirming such findings (e.g., \citealt{2007ApJ...660L..35K, 2011MNRAS.416.1936T, 2012A&A...546A.118V}).  Recent lensing work at high redshift has also found evidence for scatter to low rotation velocity from the TF relation for a handful of low mass galaxies \citep{2010MNRAS.404.1247J, 2015MNRAS.450.1812L}. However, due to a combination of sensitivity limitations and survey selection, the low mass regime of the high-redshift TF relation has not been explored with large numbers of galaxies. We will now explore the low mass TF relation at $z\sim2$ with SIGMA galaxies.

\subsection{TF Relation for SIGMA}
\begin{figure}
\begin{centering}
\includegraphics[angle=0,scale=.38]{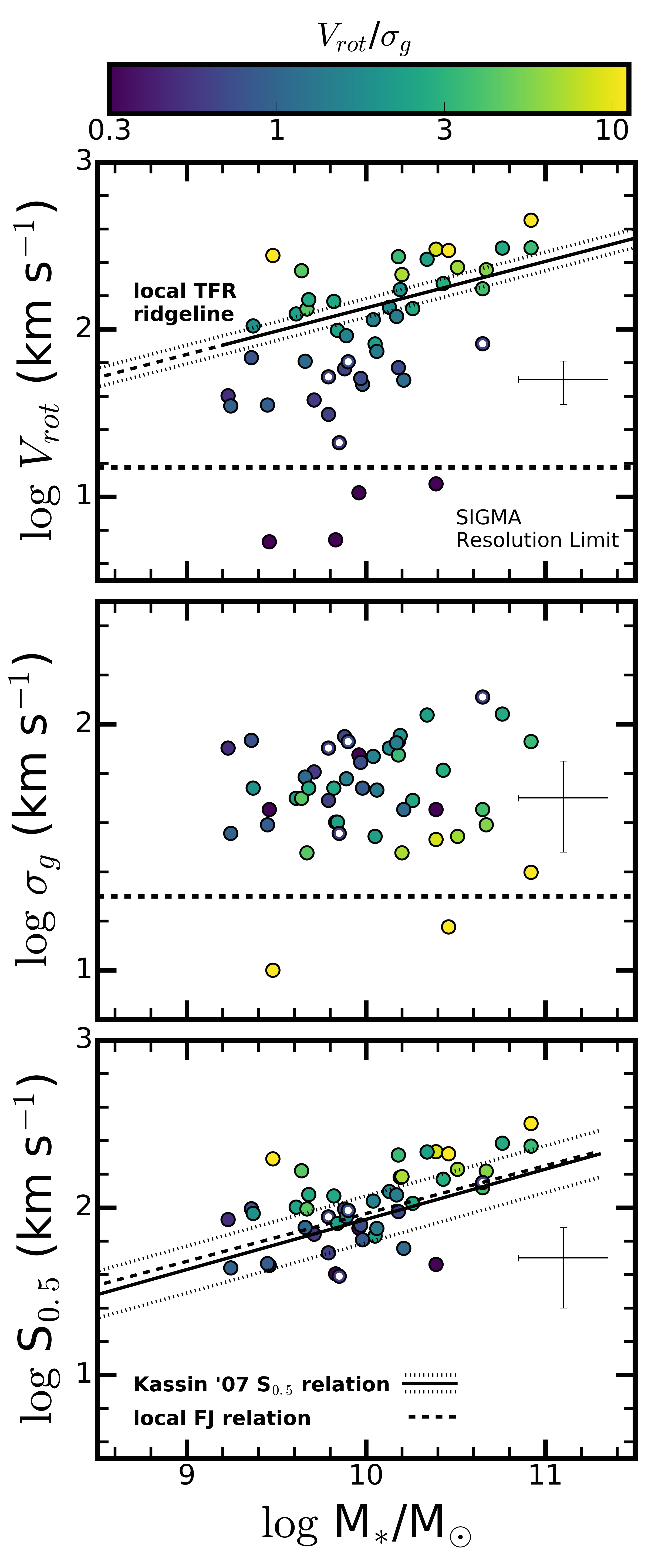}
\caption {Top: The Tully-Fisher (TF) relation for SIGMA shows significant scatter to low $V_{rot}$ from the local TF ridge line, defined by \citealt{2011MNRAS.417.2347R}. The largest scatter is found at low stellar mass ($\log\,M_{*}/M_{\odot}\,\lesssim\,10.2$). Middle: The relation between integrated gas velocity dispersion ($\sigma_g$) and $M_*$ is shown. The slowly rotating galaxies that fall from the TF relation have elevated velocity dispersions. Bottom: The S$_{0.5}$ TF relation is shown along with the fit and scatter from the $z\sim0.2$ DEEP2 sample \citep{2007ApJ...660L..35K} and the local Faber-Jackson relation, defined by \citealt{2006MNRAS.370.1106G}. The handful of galaxies in which the largest velocity gradient was measured in an off-axis slit are not inclination corrected and are shown as open circles. The resolution limits for measuring rotation and dispersion in SIGMA is given by the horizontal dashed lines in the top and middle plots.}
\label{fig:kinematics}
\end{centering}
\end{figure}

In the top panel of Figure \ref{fig:kinematics} we plot the stellar mass Tully-Fisher relation for the galaxies in SIGMA. The points are color coded to indicate the ratio of ordered ($V_{rot}$) to disordered ($\sigma_g$) motions in each galaxy, which we will refer to as ``kinematic order". Only one PA, the PA with the strongest velocity gradient, is shown for each galaxy. As described in \S 4.1, we do not apply an inclination correction for the 4 galaxies whose largest velocity gradients are measured at a $\Delta$ PA $>$ 45$^{\circ}$. A few galaxies have a best-fit rotation velocity which is below the nominal resolution limit of our spectra, shown by the dashed line. We set these galaxies at an upper limit of $V_{rot}$ $\times$ $\sin i$ $<$ 15 km s$^{-1}$ in subsequent figures.

The most massive galaxies in SIGMA ($\log M_*/M_{\odot}\,\gtrsim\,10.2$) lie on or above the TF ridge line, which we define using the local relation from \citealt{2011MNRAS.417.2347R}. These galaxies exhibit values of $V_{rot}/\sigma_g\,\sim2-8$, factors of a few lower than local galaxies at the same mass. We perform a simple linear regression to the stellar mass TF relation for these 12 massive ($\log M_*/M_{\odot}\,\gtrsim\,10.2$) rotationally-supported galaxies ($V_{rot}/\sigma_g\,>\,1$). We fix the slope to the value from local relation and report an offset to lower $M_*$ of $0.44\,\pm\,0.16$ $dex$. If these systems are gas-rich, then the baryonic ($M_*$+$M_{gas}$) TF relation is the more applicable scaling relation. Although measurements of gas mass are unavailable for the galaxies in our sample, we note that an offset in mass of $0.44\,dex$ implies that we may be missing $\sim\,60\%$ of the baryonic mass. This is consistent with estimates of the molecular gas reservoirs in massive galaxies at high redshift (e.g., \citealt{2013ApJ...768...74T}). As SIGMA includes both rotationally-supported and dispersion-dominated galaxies, we make no further attempt to characterize an evolution in the slope, scatter or intercept of the stellar mass TF relation for disks, which would require morphological pruning. At low stellar stellar mass ($\log\, M_*/M_{\odot}\,<\,10.2$) we find significant scatter from the ridge line to low rotation velocity. A large fraction of these galaxies show marginal rotation support, with values of $V_{rot}/\sigma_g\,\lesssim\,1$.

In the middle panel of Figure \ref{fig:kinematics} we show the velocity dispersion as a function of stellar mass. As discussed in \S 3, the velocity dispersion that we measure is corrected for seeing and modeled as a constant term across the galaxy. Due to the finite width of the slit (0.7$\arcsec$) and the effects of seeing ($\sim$ 0.6$\arcsec$), $\sigma_g$ is an integration of small scale velocity gradients, thermal broadening and internal turbulence. The velocity dispersion is only mildly correlated with stellar mass, with more massive galaxies displaying marginally higher velocity dispersions. The mean and rms scatter of $\sigma_g$ for the low mass ($9\,<\,\log\,M_*/M_{\odot}\,<\,10$) and high mass ($10\,<\,\log\,M_*/M_{\odot}\,<\,11$) samples are 57$\,\pm\,$19 km s$^{-1}$ and 64$\,\pm\,$29  km s$^{-1}$, respectively. 
The galaxies which fall from the TF ridgeline to low $V_{rot}$ display characteristically higher $\sigma_g$ at fixed stellar mass. In other words, the lower $V_{rot}$/$\sigma_g$ in these galaxies is due to both lower rotation velocity and higher velocity dispersion.

Star-forming galaxies at z $\sim$ 2 tend to have much stronger kinematic contributions from $\sigma_g$ than their local counterparts. For the bulk of our sample there appears to be a floor in velocity dispersion of $\sigma_g$ = 30 km s$^{-1}$. This floor is slightly higher than the typical gas velocity dispersion in local disk galaxies, $\sim\,25$ km s$^{-1}$ \citep{2010MNRAS.401.2113E}, and higher than the spectral resolution limit of our observations, $\sim$ 20 km s$^{-1}$.

To combine the quantities $V_{rot}$ and $\sigma_g$ into a more fundamental quantity, many authors (e.g., \citealt{2006ApJ...653.1049W, 2007ApJ...660L..35K, 2012A&A...546A.118V}) adopt a joint kinematic parameter, $S_K = \sqrt{K\,V_{rot}^2 + \sigma_g^2}$ \citep{2006ApJ...653.1049W}. A common choice is made for $K\,=\,0.5$, motivated by virial arguments for a spherically symmetric tracer distribution (see \citealt{2006ApJ...653.1027W}).  The $S_{0.5}$ quantity more reliably traces the overall potential of a galaxy system, independent of morphology, recent disturbance or kinematic state \citep{2007ApJ...660L..35K, 2010ApJ...710..279C}. In the bottom panel of Figure \ref{fig:kinematics} we show the $M_*\,-\,S_{0.5}$ relation for our sample, which re-establishes the TF relation for all galaxies. This has been found previously at $z\sim2$ \citep{2009ApJ...697..115C, 2016ApJ...819...80P}. We note that our low mass galaxies are relatively consistent with the $z\sim0.2$ relation found by \citet{2007ApJ...660L..35K}, while the high mass galaxies tend to be offset to higher $S_{0.5}$. \citet{2016ApJ...819...80P} report a similar offset in the $z\sim2$ $S_{0.5}$ relation over their full mass range ($9\,\lesssim\,\log\,M_*/M_{\odot}\,\lesssim\,11$). \citet{2016arXiv160204813R} have argued that it is necessary for the $M_*-S_{0.5}$ to evolve to higher $S_{0.5}$ at fixed mass in order to correctly predict the cosmic stellar mass density. As discussed before, this offset might reflect a rise in $M_{gas}/M_{baryon}$ at higher redshift.

\begin{figure*}
\begin{centering}
\includegraphics[angle=0,scale=.33]{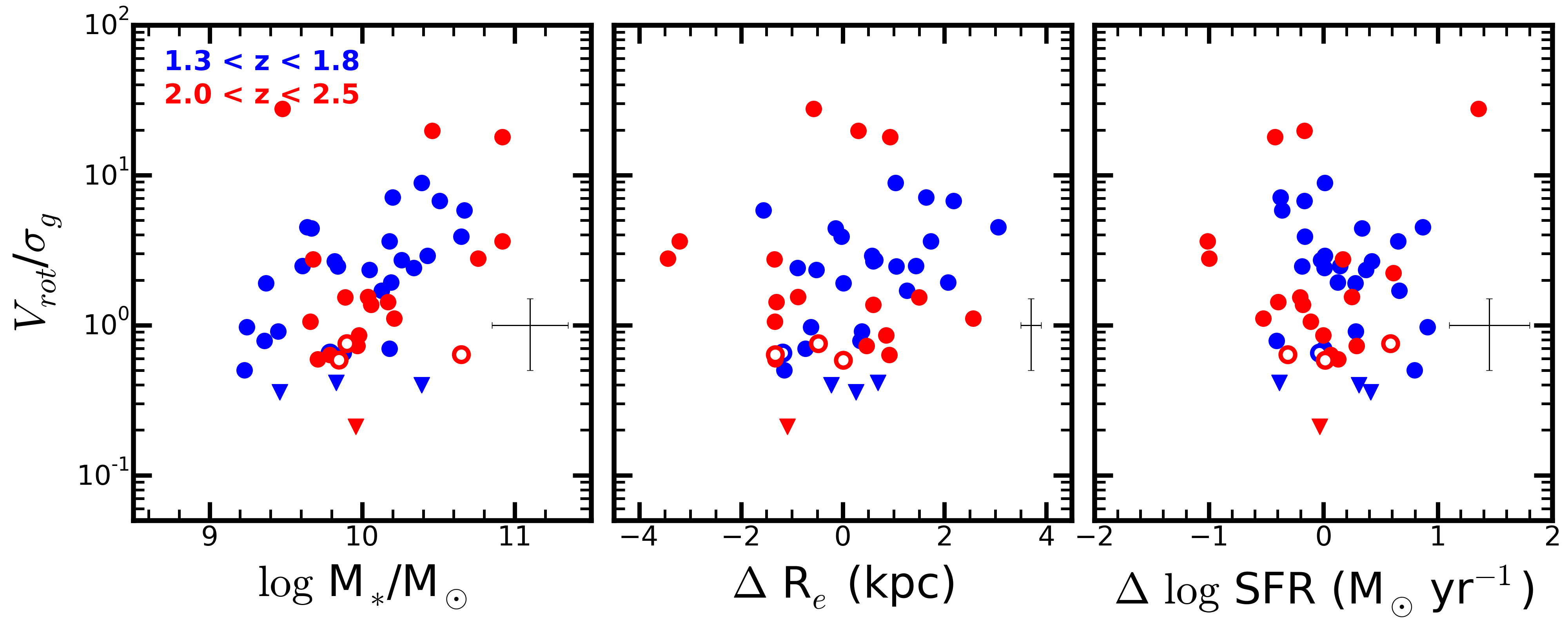}
\caption {The kinematic order ($V_{rot}/\sigma_g$) of SIGMA galaxies is strongly correlated with stellar mass ($M_*$; left panel), moderately correlated with H-band continuum size at fixed a stellar mass ($\Delta\,R_e$; middle panel) and weakly correlated with with star-formation rate at fixed stellar mass ($\Delta \log$ SFR; right panel). In the 3 panels we show $V_{rot}/\sigma_g$ versus $M_*$, residuals from the $z\sim2$ SFR-$M_*$ relation and residuals from the $z\sim1.75$ and $z\sim2.25$ size-mass relations. The blue and red points represent our low and high redshift subsamples, respectively. Typical error bars for each quantity are shown in each plot. Galaxies with measured values of $V_{rot}$ below the spectral resolution limit are set to the resolution limit and are shown as upper-limit symbols (upside-down triangles).}
\label{fig:vsig_others}
\end{centering}
\end{figure*}

We perform a least-squares linear fit to the $S_{0.5}$ TF relation, $\log$ $S_{0.5}$ $($km s$^{-1})$ = $\alpha\,(\log M_*/M_{\odot} - 10) + \beta$, for the full sample. We report coefficients of $\alpha=0.25\,\pm\,0.06$ and $\beta = 2.00 \,\pm\, 0.03$ with an rms scatter $\sigma_{rms}$ = 0.19. Even with the apparent dichotomy between the low mass and high mass regions noted above, we find that the slope, amplitude and scatter of the full sample is relatively consistent with the $z\sim0.2$ relation ($\alpha = 0.30$, $\beta = 1.93$, $\sigma_{rms}$ = 0.14) found in \citet{2007ApJ...660L..35K} and the local Faber-Jackson relation ($\alpha$ = 0.286, $\beta$ = 1.965)\citep{2006MNRAS.370.1106G}.

\section{Kinematic Order vs star-formation rate, size and mass}

As discussed in \S 5, the low mass galaxies in our sample show the strongest residuals from the TF ridge line and the lowest values of $V_{rot}/\sigma_g$. In the settling of disk galaxies, massive galaxies are expected to form ordered disks first, followed later by lower mass galaxies (``kinematic downsizing"; \citealt{2012ApJ...758..106K}). To probe the drivers of kinematic disturbance at $z\sim2$, we examine the relation between the kinematic order of SIGMA galaxies ($V_{rot}$/$\sigma_g$) and three of their physical properties: star-formation rate, effective radius and stellar mass. To separate out the dependance of size and SFR on stellar mass, we examine deviations from the size-mass and SFR-mass relations at $z\sim2$. We find that, at fixed stellar mass, $V_{rot}$/$\sigma_g$ is only weakly correlated with SFR and mildly correlated with half-light radius. The strongest correlation is found to be with stellar mass, wherein more massive galaxies tend to be the most rotationally supported. We describe these relations in more detail below.

We perform a ranked spearman correlation test to quantify the association in each relation. The spearman coefficient ($r_s$) measures the strength of a correlation and we use the common interpretation: $|r_s|\,<\,0.15$ (very weak correlation), $0.15 <\,|r_s|\,<\,0.30$ (weak/moderate correlation), $0.30 <\,|r_s|\,<\,0.50$ (moderate/strong correlation). The p-value ($p$) indicates the probability of uncorrelated data producing the given spearman coefficient.

In the left panel of Figure \ref{fig:vsig_others} we examine the correlation of $V_{rot}/\sigma_g$ with stellar mass following K12 for $z\,<\,$1.2. We note that the trend is the strongest of the three we investigate, with an increasing trend of $V_{rot}$/$\sigma_g$ with stellar mass and a spearman coefficient, $r_s = 0.39$ ($p\,=\,0.01$). If we only include the galaxies with detected rotation and inclination corrections (i.e., filled circles), the trend with stellar mass strengthens to $r_s = 0.48$ ($p\,<\,10^{-2}$).

Next, we examine the departure of our galaxies from the size-mass relation at $z\sim2$, as parameterized for late type galaxies in Table 1 of \citealt{2014ApJ...788...28V}. We use the $z\sim1.75$ and $z\sim2.25$ size-mass relations for our low and high redshift bins, respectively. The half-light size used for these relations is calibrated to rest-frame 5000 \AA\,(1.5 $\mu$m at z$=$2) and so we adopt the H-band half-light radii for comparison. In the middle panel of Figure \ref{fig:vsig_others}, we present the differences between the half-light radii of our galaxies and the size-mass relation ($\Delta\,$R$_e$ = R$_{e,gal}$ - R$_{e,seq}$.). The intrinsic scatter of the relation is 0.18 and 0.19 dex for both the low and high redshift bins, respectively, consistent with the rms scatter in our sample of 0.19. We note that there is a weak/mild correlation between $V_{rot}$/$\sigma_g$ and $\Delta$ R$_e$, as quantified by the spearman coefficient $r_s\,= 0.23$ ($p\,=\,0.10$).

Finally, in the right panel of Figure \ref{fig:vsig_others}, we examine the correlation of kinematic order with a galaxy's departure from the 1.5 $<$ z $<$ 2.6 star-formation main sequence ($M_*-SFR$). If the dispersion-dominated galaxies are being disrupted by internal processes associated with star-formation, we would expect a correlation between $V_{rot}/\sigma_g$ and the position of a galaxy on the star-formation main sequence. We adopt the main sequence fit of \citet{2012ApJ...754...25R} as parameterized in \citet{2014ApJS..214...15S} and adjust to a Chabrier IMF to match with our sample. This relation is valid over $\log\,M_*/M_{\odot}\,=\,8.7\,-\,11.1$, spanning our full mass interval. We find no strong trend with $\Delta\log$ SFR (defined as $\log$ SFR$_{gal}$ - $\log$ SFR$_{seq}$) and $V_{rot}$/$\sigma_g$. In other words, at fixed stellar mass the position of a galaxy above or below the star-formation main sequence is not related to its internal gas kinematics. The spearman coefficient between these two values confirms the lack of a trend, $r_s\,= -0.09$ (with $p\,=\,0.54$).

\begin{figure*}
\begin{centering}
\includegraphics[angle=0,scale=.65]{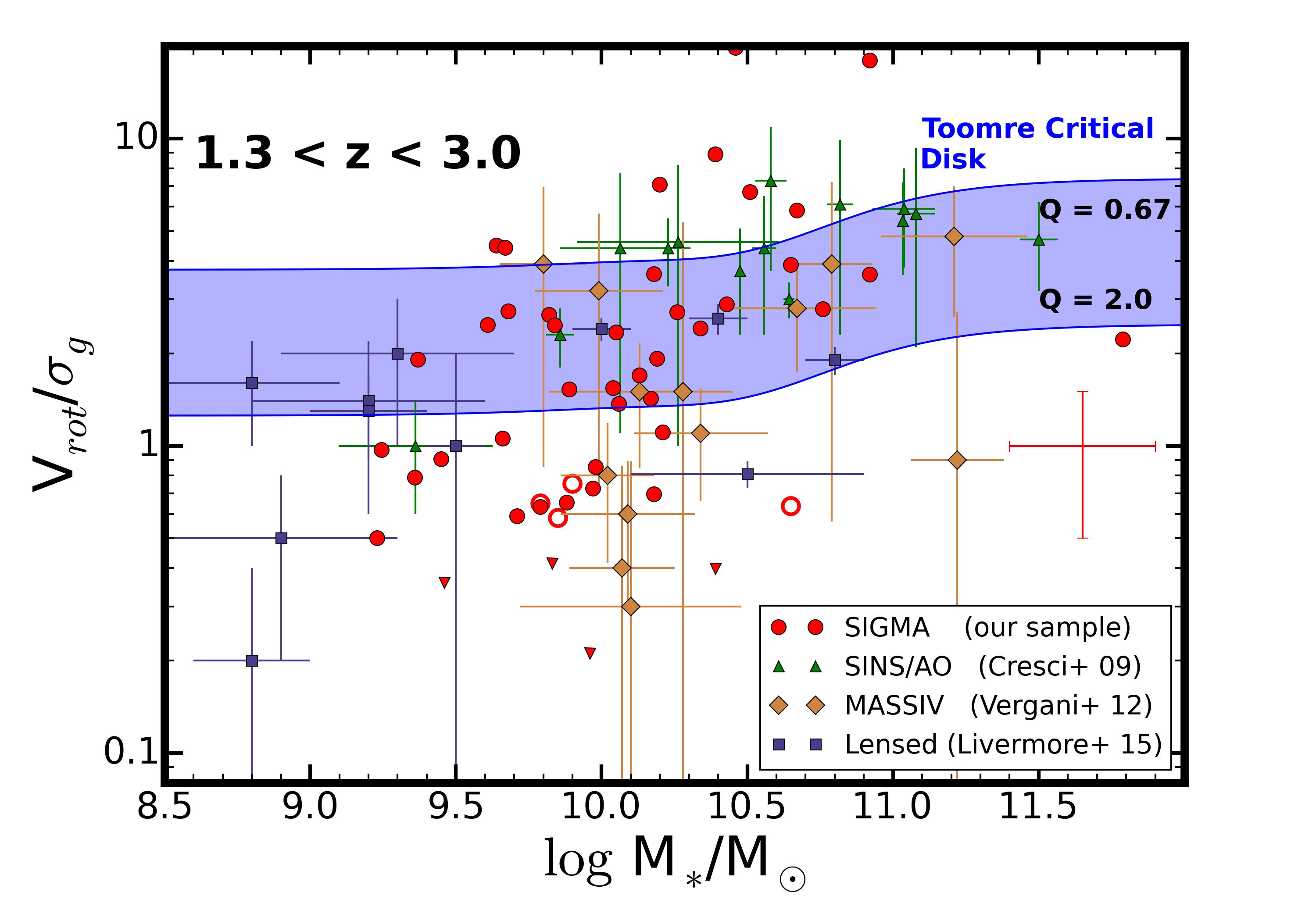}
\caption {Kinematic order ($V_{rot}$/$\sigma_g$) versus stellar mass for galaxies in SIGMA and from the literature. A model for a marginally stable disk galaxy at $z\sim2$ is shown as a blue shaded swath. While high mass galaxies fall within this swath, a large fraction of the low mass galaxies fall below it. High mass galaxies at $z\sim2$ are consistent with the model, implying that they have already formed primitive disks. In contrast, a large fraction of the low mass galaxies are dispersion dominated ($V_{rot}/\sigma_g\,<\,1$) and are still in the process of assembling their disks. We note that the high mass disks are unlike the majority of local disk galaxies at a similar stellar mass, which typically have values of $V_{rot}$/$\sigma_g$ of around 10-15.}
\label{fig:vsig_m}
\end{centering}
\end{figure*}

Combining kinematic measurements at $z\sim2$ from both the KMOS-3D \citep{2015ApJ...799..209W} and MOSDEF surveys \citep{2015ApJS..218...15K}, \citealt{2016ApJ...819...80P} investigated the trend of V/$\sigma_g$ with specific star formation rate ($sSFR = SFR/M_{*}$) and stellar mass, finding a declining and weakly increasing trend respectively. We find that, when we fix for stellar mass, there is no dependance on SFR. Instead, the strongest correlation in the SIGMA sample is with stellar mass alone. If we consider that the gas mass scales with SFR \citep{1998ApJ...498..541K, 2010ApJ...714L.118D}, these results would indicate that the kinematic order of a galaxy is more strongly linked with its mass than its gas fraction.

\section{A Test for Stable Disks}

In this section we compare our measurements of $V_{rot}$/$\sigma_g$ with a model for gas rich marginally stable disks using the refined Toomre analysis given in \citet{2011ApJ...733..101G}. We find that high mass galaxies in our sample are in general agreement with this model, indicating that they have already assembled primitive disks. 
 
 \subsection{Model for a stable disk}
Galaxies which have already settled to become disks by $z\sim2$ are gas rich (e.g., \citealt{2010ApJ...713..686D, 2013ApJ...768...74T}) and are expected to be in a marginally stable (to collapse) equilibrium (e.g., \citealt{2008ApJ...687...59G}), regulated by ubiquitous inflows, outflows and star-formation \citep{2010ApJ...718.1001B, 2012MNRAS.421...98D}. This regulator model successfully describes the elevated velocity dispersions in high redshift gas-rich disks \citep{2015ApJ...799..209W}.  Following the work of \citet{2011ApJ...733..101G} and \citet{2015ApJ...799..209W}, $V_{rot}/\sigma_g$ of the ionized gas in a marginally stable disk can be estimated from the Toomre criterion:

\begin{equation}
\frac{V_{rot}}{\sigma_g} = \left(\frac{a}{Q_{crit}}\right)\left(\frac{1}{f_{gas}}\right)
\end{equation}                                               
where $a$ is a constant which is set to $\sqrt2$ for a flat rotation curve, as discussed in \citet{2009ApJ...703..785D}. The quantity $Q_{crit}$ is the critical Toomre parameter which, considering the range of realistic disks, can vary between 0.67 (single-component thick disk; \citealt{2011ApJ...733..101G}) and 2.0 (two-component thin disk; \citealt{2010MNRAS.404.2151C}). The gas fraction ($f_{gas}$) can be estimated from the gas depletion timescale ($t_{depl} = M_{mol,gas}/SFR$) and the specific star-formation rate ($sSFR$) as $1/f_{gas}=1+(t_{depl}\times sSFR)^{-1}$. The gas depletion timescale is relatively constant with stellar mass at $z\sim2$ and we fix it to $t_{depl} = 7\times10^8$ yr (\citealt{2013ApJ...768...74T, 2015ApJ...800...20G}). For our analysis we adopt the sSFR - M$_*$ relation at z = 2 from \citet{2014ApJ...795..104W} using the fit in \citet{2015ApJ...799..209W}. Equation 2 is derived assuming that the gravitational force exerted by $M_{dyn}$ is supported by rotation, which is not generally true for our dispersion dominated systems. This stability analysis would not directly apply to these galaxies, but we can still compare $V_{rot}/\sigma_g$ in these systems with our disk model.

\subsection{Applying the model to data}
In Figure \ref{fig:vsig_m} we show $V_{rot}$/$\sigma_g$ as a function of stellar mass for the galaxies in our sample and additional publicly available measurements for the following surveys:  SINS \citep{2009ApJ...697..115C}, MASSIV \citep{2012A&A...546A.118V}, and lensed galaxies from \citet{2015MNRAS.450.1812L} and \citet {2010MNRAS.404.1247J}. The SINS sample is the high signal-to-noise adaptive optics assisted subset and is representative of the most massive and star-forming galaxies at $z\sim2$. The lensed sample and MASSIV sample are more generally representative of star-forming galaxies at these epochs. We use these samples to extend our mass baseline to $\log M_*/M_{\odot} = 8\,-\,11.5$ and to illustrate the consistency of our measurements with those in the literature.

The expected distribution of $V_{rot}$/$\sigma_g$ for marginally stable disks at $z\sim2$ is shown in Figure \ref{fig:vsig_m} and follows from the model described above. This range includes a characteristic break above $\log M_*/M_{\odot} \approx 10.2$, due to a kink in the $z\sim2$ stellar mass sSFR relation.

We find that high mass galaxies ($\log\,M_*/M_{\odot}\,>\,10.2$) are in general agreement with the marginally stable disk model, i.e., high mass galaxies have already assembled their disks by $z\sim2$. In general these disks are expected to contain large gas reservoirs which drive violent disk instabilities and maintain elevated velocity dispersions. These galaxies are unlike local disks of the same mass, which have values of $V_{rot}/\sigma_g\,\sim\,10-15$ as opposed to $V_{rot}/\sigma_g\,\sim\,2-8$. As their gas reservoirs deplete, they are expected to stabilize and form well-ordered disks \citep{2015ApJ...799..209W}.

Below a stellar mass of $\log\,M_*/M_{\odot}\,\approx\,10.2$ we find the common occurrence of systems which are dispersion dominated and/or marginally rotationally supported ($V_{rot}$/$\sigma_g$ $\lesssim$ 1). For comparison, the most dispersion dominated system expected from the disk model is $V_{rot}$/$\sigma_g$ = 0.7 (f$_{gas}$ = 1, Q$_{crit}$ = 2 in Eq 2). Galaxies with lower $V_{rot}$/$\sigma_g$ would be considered globally stable to gas collapse (Q $>$ Q$_{crit}$), even as they are obviously forming stars. Although gas-rich disks with high velocity dispersion are expected at $z\sim2$, a large fraction of our low mass sample cannot be described by this simple model for marginally stable disks.

\section{Conclusions}

We present the first results of SIGMA (Survey in the near-Infrared of Galaxies with Multiple position Angles), a kinematic survey of 49 star-forming galaxies over 1.3 $<$ z $<$ 3.0 in three of the CANDELS fields. SIGMA is representative of the $z\sim2$ star-formation main sequence to $\log\,M_*/M_{\odot}$ = 9.2 at $z\sim1.6$ and $\log\,M_*/M_{\odot}$ = 9.6 at $z\sim2.3$, and includes galaxies with both disk and irregular morphologies. Internal kinematics (rotation velocity $V_{rot}$ and gas velocity dispersion $\sigma_g$) are measured from resolved rest-optical nebular emission lines (H$\alpha$, \OIII$\lambda$ 5007) in Keck/MOSFIRE NIR spectra. 

We find that high mass galaxies ($\log\,M_*/M_{\odot}> 10.2$) at $z\sim2$ are generally rotationally supported, fall on the TF relation, and have formed primitive disks with high $\sigma_g$. These marginally stable disks are expected to decline in $\sigma_g$ with time as they deplete their gas supply \citep{2015ApJ...799..209W}. In contrast,  a large fraction of low mass galaxies ($\log\,M_*/M_{\odot}< 10.2$) at this epoch are in the early phases of assembling their disks. They have much lower ``kinematic order" ($V_{rot}\lesssim\sigma_g$) and tend to fall to low $V_{rot}$ from the Tully-Fisher relation. For perspective, these results imply that a Milky-Way progenitor at $z\sim2$ ($\log\,M_*/M_{\odot}\approx9.8$; \citealt{2015ApJ...803...26P}) is likely still in the process of forming a disk. Combining both $V_{rot}$ and $\sigma_g$ in $S_{0.5}$, all of the galaxies in SIGMA fall on the $S_{0.5}$ Tully-Fisher relation, indicating that many galaxies at low mass have strong contributions of kinematic support in the form of $\sigma_g$. Following \citet{2012ApJ...758..106K} at $z\,<\,$1, the tendency for high mass galaxies to develop disks first is referred to as ``kinematic downsizing", and these results imply that it is in place at $z\sim2$. 

To gain insight into the underlying drivers of kinematic disruption, we explore the dependance of $V_{rot}$/$\sigma_g$ on three physical parameters of galaxies: size, star-formation rate and stellar mass. We find that the strongest trend is with stellar mass, wherein high mass galaxies have the strongest rotational support (i.e., highest values of $V_{rot}/\sigma_g$). We examine the relation between $V_{rot}/\sigma_g$ and residuals from the star-formation main sequence and we find no correlation, to within the uncertainties of our measurements. Given that the star-formation rate is likely tracing the available gas mass \citep{2010ApJ...714L.118D}, our results imply that mass may be more fundamental than gas fraction in determining kinematic order. Finally, we find a moderate correlation between $V_{rot}/\sigma_g$ and size, likely associated with the increased angular momentum of the rotationally supported systems.  The strong trend between $V_{rot}/\sigma_g$ and mass and the lack of a correlation with SFR at fixed mass implies that the processes that maintain $\sigma_g$ in galaxies should be most efficient at low stellar mass and should be independent on gas fraction. 

Several processes which disrupt disk galaxies are active at $z\sim2$, associated with elevated star-formation \citep{2014ARA&A..52..415M} and rising merger rates \citep{2015MNRAS.449...49R}. Several of these mechanisms may play a role in preferentially stalling disk formation in low mass galaxies. Feedback through stellar winds and supernovae can effectively blow out gas and dynamically heat stars in low mass dwarf galaxies ($\lesssim\,10^{9.5}\,M_{\odot}$; e.g, \citealt{2015arXiv151101095W, 2016ApJ...820..131E}). We note that we find no trend between $V_{rot}/\sigma_g$ and $SFR$ at fixed mass among the galaxies studied in this sample ($9.2\,<\,\log M_*/M_{\odot}\,<\,11$), indicating that such feedback mechanisms may only assume a secondary role in disturbing the gas kinematics in these galaxies. Mergers play a complicated role in shaping the angular momentum content of disk galaxies. Models (e.g., \citealt{2009MNRAS.398..312G, 2009ApJ...691.1168H}) and observations (e.g., \citealt{2005A&A...430..115H, 2009A&A...507.1313H}) indicate that gas-rich disks can survive major mergers, albeit with thick disks and perturbed kinematics.  Examining mock observations from a suite of simulated galaxy mergers, \citet{2010ApJ...710..279C} find that merger remnants tend to scatter to low $V_{rot}$ from the TFR, similar to the low mass galaxies in our sample. Although several channels exist for stalling (and promoting) disk formation, it is as of yet unclear what the dominate processes are in maintaining the elevated kinematic disturbance we observe in low mass galaxies at this epoch.

In summary, we find that kinematic downsizing is active as far back in time as $z\sim2$. While massive star-forming galaxies ($\log\,M_*>10.2$) appear to have assembled primitive rotating disks by this time (consistent with e.g., \citealt{2006Natur.442..786G, 2006ApJ...645.1062F, 2008ApJ...687...59G, 2008ApJ...682..231S, 2009ApJ...706.1364F, 2011ApJ...742...11S, 2012Natur.487..338L, 2015ApJ...799..209W, 2016ApJ...819...80P}), low mass galaxies tend to be more kinematically disordered and still in the early process of assembling their disks. The peak of cosmic star-formation is coincident with the epoch of disk assembly for low mass galaxies and the settling of primitive disks in high mass galaxies.

\section*{Acknowledgements}

The authors would like to thank the anonymous referee who provided useful suggestions that improved this paper. RCS would like to thank Sandra Faber, Joel Primack, Avishai Dekel and the CANDELS-Baltimore group for useful comments and discussions which have also improved the paper. RCS and SAK gratefully acknowledge the support from a grant through the STScI JDF. We thank G. Wirth and the TKRS-2 team for their enormous effort observing, reducing and releasing the MOSFIRE spectra which are used for the SIGMA galaxies in GOODS-N. We wish to extend thanks to those of Hawaiian ancestry on whose sacred mountain we are privileged guests. This work has made use of the Rainbow Cosmological Surveys Database, which is operated by the Universidad Complutense de Madrid (UCM), partnered with the University of California Observatories at Santa Cruz (UCO/Lick,UCSC). JRT acknowledges support from NASA through Hubble Fellowship grant $\#$51330 awarded by the Space Telescope Science Institute, which is operated by the Association of Universities for Research in Astronomy, Inc., for NASA under contract NAS 5-26555. CP is supported by an appointment to the NASA Postdoctoral Program at NASA's Goddard Space Flight Center.

\newpage

\begin{figure*}
\begin{centering}
\includegraphics[angle=0,scale=.40]{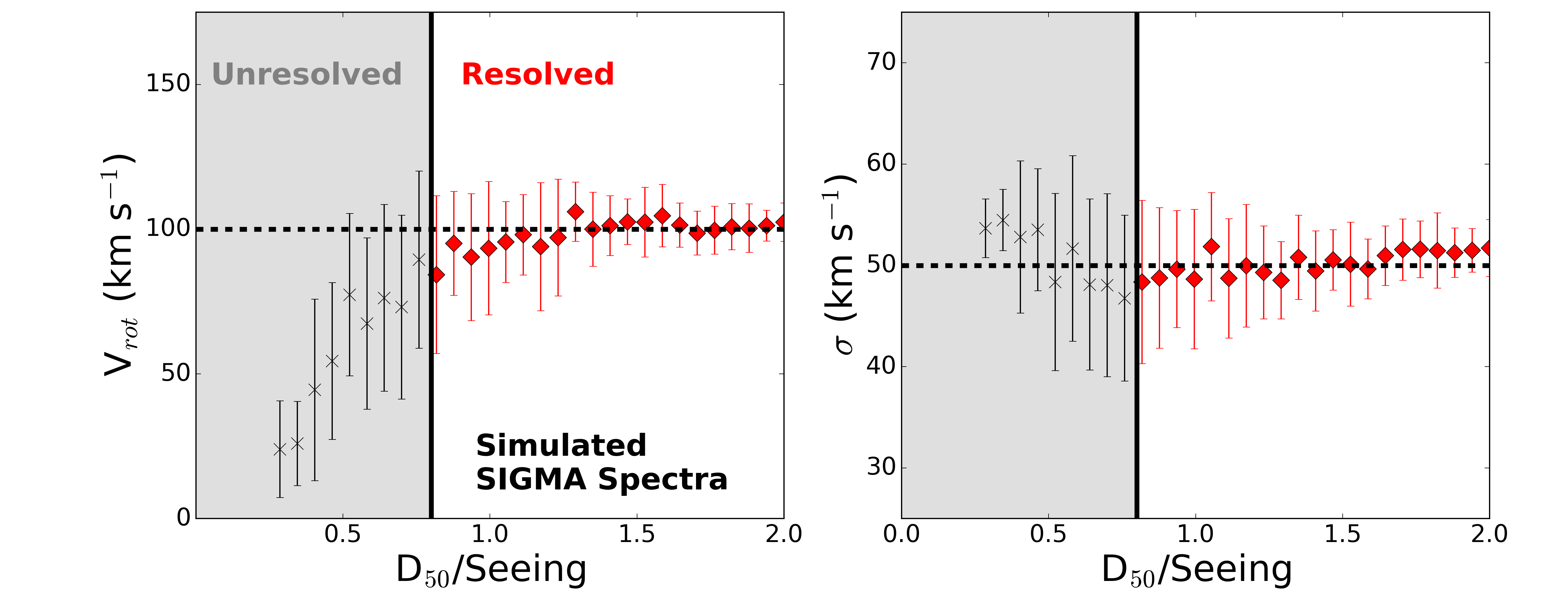}
\caption {Our models demonstrate that galaxy kinematics in SIGMA can be measured down to a size of $D_{50} = 0.8\,\times\,Seeing_{FWHM}$, where $D_{50}$ is the intrinsic half-light diameter and $Seeing_{FWHM}$ is the full width at half maximum of the seeing during the observations. These plots show mock observations of a model galaxy at varying intrinsic sizes, the points show the mean and rms scatter from 100 realizations at each size interval. The model is a rotation dominated galaxy with $V_{rot}\,=\,100 $ km s$^{-1}$ and $\sigma_g\,=\,50$ km s$^{-1}$, marked by the dashed line, and the emission line profile is a Sersic with an index of $n\,=\,2$. The correct values of $V_{rot}$ (left panel) and $\sigma_g$ (right panel) are recovered (red circles) for mock observations with intrinsic sizes $D_{50}>0.8\,\times\,Seeing_{FWHM}$. We find a systematic drop in $V_{rot}$ and a slight increase in $\sigma_g$ for models that have sizes smaller than this value and are unresolved (black x's).}
\label{fig:model_size}
\end{centering}
\end{figure*}

\appendix
\section{Modeling the effects of size}

In this Appendix, we perform mock observations of simulated SIGMA spectra in order to determine the limiting spatial scale from which we can recover rotation velocity, if present, in our observations. We find that we can recover kinematics down to a limit of D$_{50}$/Seeing$_{FWHM}$ = 0.8.  In \S 2, we apply this limit to our data set.

The spectra presented in this paper were observed in seeing limited conditions, and as such the effects of beam smearing \citep{1987PhDT.......199B} are important to understand. Seeing tends to blur together spatially distinct velocities, leading to a decreased rotation signature and an elevated velocity dispersion. To explore the effective resolution set by beam smearing in SIGMA, we create a set of model spectra with various sizes and fixed kinematic parameters. As described below, we perform mock observations of the kinematics for these spectra and explore how the results vary with size. Our model for a single mock SIGMA/MOSFIRE spectrum consists of first creating a 1D emission line with a Sersic intensity profile:

\begin{equation}
\Sigma(r) = \Sigma_e e^{-k [(r/R_e)^{1/n}-1]}
\end{equation}
\noindent
where R$_e$ is the half-light radius, $\Sigma_e$ is the intensity at the half-light radius, $n$ is the sersic index and $k$ is a numerical constant that depends on $n$ ($k = 2n-0.3271$). We set $n\,=\,1.8$, characteristic of the emission line intensity profile for galaxies at this redshift \citep{2015arXiv150703999N}. We then draw from an appropriate error spectrum to set the central signal-to-noise per pixel to 15, typical of SIGMA spectra. The final integrated S/N of these simulated spectra are well-matched to our sample. Next, we apply an arctan velocity profile (see Eq 1) and a constant velocity dispersion to the emission line. We fix the rotation velocity to 100 km s$^{-1}$ and the velocity dispersion to 50 km s$^{-1}$, characteristic of $z\sim2$ disk galaxies. We then apply a seeing kernel of 0.7" in the spatial direction and MOSFIRE instrumental broadening ($\sigma_{inst}$ = 40 km s$^{-1}$) in the spectral direction. Following the seeing convolution, the emission profile is well described by a gaussian ($n\,=\,0.5$). We repeat this process 100 times, redrawing from the error spectrum each time to create independent realizations with the same fixed parameters. We then vary the effective radius of the emission line from 0.1 - 0.7 $\arcsec$ in 30 even steps of 0.02$\arcsec$/step. At each step we again simulate 100 realizations with the same fixed properties. The final dataset contains 3000 simulated SIGMA spectra.

To measure the kinematics from the mock spectra, we run the same fitting technique used for our observations ({\tt ROTCURVE}). For each spatial scale modeled, we measure 100 values of $V_{rot}$ and $\sigma_g$ and examine their mean and variance. The results of this study are shown in Figure \ref{fig:model_size}. The rotation velocity is well recovered with an intrinsic error of $\sim$ 15 km s$^{-1}$.  For simulated spectra with effective diameter at or greater than the seeing, we accurately recover the input rotation and dispersion. Below a value of D$_{50}$/Seeing$_{FWHM}$ = 0.8, the limiting spatial resolution drastically affects the recovered values of $V_{rot}$ and $\sigma_g$. There is both a larger uncertainty in the measurements and a systematic decline in the measured rotation velocity. Based on the results of these models, we determine that the spatial scale for recovering kinematics in our observations is diameter D$_{50}$ $>$ 0.8 $\times$ Seeing$_{FWHM}$. Above this value we can reliably recover rotational motion, if present.

\newpage

\begin{deluxetable*}{ccllccccc}
\tabletypesize{\footnotesize} 
\tablecolumns{13} 
\tablewidth{0pt} 
\tablecaption{SIGMA - Observational and Physical Properties}
\tablehead{ 
\colhead{ID} & & \colhead{RA (J2000)} & \colhead{DEC (J2000)} & \colhead{z$_{spec}$}&  \colhead{m$_H$ (AB)} &\colhead{$\log M_{*} \,(M_{\odot})$} & \colhead{$\log SFR \,(M_{\odot} \,yr^{-1})$} & \colhead{$(b/a)_H$}}
\vspace{0.3cm}
\startdata 
5492  & & 03 32 41.78 & $-$27 51 35.28 & 1.67 & 22.51 & 10.39 & 1.95 & 0.57 \\
14602 & & 03 32 14.66 & $-$27 47 02.70 & 1.73 & 22.37 & 10.18 & 1.44 & 0.91 \\
16209 & & 03 32 15.82 & $-$27 46 22.36 & 1.33 & 23.47 & 9.36 & 0.23 & 0.34 \\
16600 & & 03 32 30.71 & $-$27 46 17.17 & 1.31 & 21.20 & 10.51 & 1.59 & 0.70 \\
16905 & & 03 32 15.75 & $-$27 46 04.32 & 1.54 & 22.15 & 10.13 & 2.05 & 0.32 \\
17488 & & 03 32 20.18 & $-$27 45 49.31 & 1.61 & 23.16 & 9.46 & 1.15 & 0.41 \\
17673 & & 03 32 33.86 & $-$27 45 42.65 & 1.62 & 22.69 & 10.05 & 1.68 & 0.85 \\
18177 & & 03 32 12.48 & $-$27 45 30.42 & 1.38 & 22.03 & 10.19 & 1.57 & 0.41 \\
19818 & & 03 32 18.07 & $-$27 44 33.44 & 1.38 & 22.43 & 9.23 & 1.31 & 0.81 \\
20495 & & 03 32 11.86 & $-$27 44 13.38 & 1.33 & 21.30 & 10.43 & 1.69 & 0.78 \\
20569 & & 03 32 29.19 & $-$27 44 13.21 & 1.47 & 23.89 & 9.24 & 1.44 & 0.49 \\
20593 & & 03 32 12.77 & $-$27 44 07.77 & 1.61 & 23.13 & 10.26 & 1.49 & 0.29 \\
20883 & & 03 32 26.77 & $-$27 43 58.22 & 1.61 & 23.43 & 9.88 & 1.14 & 0.51 \\
21007 & & 03 32 31.83 & $-$27 43 56.26 & 1.55 & 21.80 & 10.67 & 1.55 & 0.81 \\
21162 & & 03 32 23.95 & $-$27 43 49.09 & 1.31 & 22.16 & 9.79 & 1.02 & 0.87 \\
21258 & & 03 32 30.75 & $-$27 43 45.34 & 1.43 & 23.17 & 9.37 & 0.93 & 0.33 \\
21737 & & 03 32 30.14 & $-$27 43 35.44 & 1.60 & 22.55 & 10.20 & 1.08 & 0.69 \\
21897 & & 03 32 34.08 & $-$27 43 28.43 & 1.60 & 22.44 & 9.82 & 1.51 & 0.42 \\
22236 & & 03 32 26.19 & $-$27 43 17.38 & 1.61 & 23.09 & 9.61 & 1.03 & 0.49 \\
22354 & & 03 32 13.66 & $-$27 43 13.08 & 1.47 & 22.45 & 9.67 & 1.28 & 0.91 \\
22783 & & 03 32 26.30 & $-$27 43 01.89 & 1.55 & 23.11 & 9.83 & 0.71 & 0.46 \\
23095 & & 03 32 38.50 & $-$27 42 27.89 & 1.61 & 21.93 & 10.39 & 1.65 & 0.76 \\
23632 & & 03 32 28.13 & $-$27 42 48.27 & 1.61 & 23.35 & 9.45 & 1.01 & 0.33 \\
25715 & & 03 32 22.72 & $-$27 41 40.74 & 1.61 & 22.12 & 10.34 & 1.60 & 0.96 \\
25973 & & 03 32 26.08 & $-$27 41 38.49 & 1.54 & 22.74 & 9.84 & 0.92 & 0.62 \\
26050 & & 03 32 13.89 & $-$27 41 58.29 & 1.32 & 22.57 & 9.64 & 1.78 & 0.24 \\
26177 & & 03 32 27.28 & $-$27 42 05.20 & 1.61 & 21.11 & 10.65 & 1.73 & 0.91 \\
26736 & & 03 32 26.41 & $-$27 42 28.30 & 1.61 & 22.21 & 10.18 & 2.09 & 0.49 \\
1826  & & 12 36 18.78 & $+$62 08 39.70 & 2.30 & 23.03 & 9.79 & 1.12 & 0.32 \\
2815  & & 12 36 35.56 & $+$62 09 20.33 & 2.36 & 23.26 & 9.71 & 1.11 & 0.66 \\
4106  & & 12 36 39.37 & $+$62 10 06.60 & 2.35 & 22.40 & 10.65 & 1.58 & 0.31 \\
4476  & & 12 36 13.13 & $+$62 10 21.11 & 2.24 & 23.50 & 9.89 & 0.95 & 0.27 \\
4714  & & 12 36 36.08 & $+$62 10 29.90 & 2.31 & 23.01 & 10.17 & 1.03 & 0.83 \\
4925  & & 12 36 11.51 & $+$62 10 33.72 & 2.25 & 22.52 & 10.92 & 1.73 & 0.72 \\
4962  & & 12 36 18.77 & $+$62 10 37.28 & 2.27 & 23.56 & 9.68 & 1.12 & 0.59 \\
9157  & & 12 36 13.56 & $+$62 12 21.58 & 2.44 & 23.13 & 9.90 & 1.75 & 0.48 \\
9190  & & 12 36 18.27 & $+$62 12 22.18 & 2.44 & 23.38 & 9.85 & 1.13 & 0.89 \\
10230 & & 12 37 06.71 & $+$62 12 44.61 & 2.21 & 23.23 & 10.21 & 0.94 & 0.17 \\
11525 & & 12 36 26.94 & $+$62 13 17.86 & 2.40 & 23.91 & 9.48 & 2.11 & 0.42 \\
13230 & & 12 36 09.05 & $+$62 13 59.03 & 2.05 & 24.83 & 9.96 & 1.19 & 0.37 \\
13678 & & 12 36 44.08 & $+$62 14 10.10 & 2.27 & 26.53 & 9.66 & 0.82 & 0.41 \\
14428 & & 12 36 35.59 & $+$62 14 24.01 & 2.02 & 25.47 & 11.79 & 3.61 & 0.86 \\
15497 & & 12 36 21.74 & $+$62 14 52.85 & 2.21 & 26.28 & 10.06 & 1.14 & 0.76 \\
16028 & & 12 36 51.82 & $+$62 15 04.72 & 2.19 & 26.38 & 10.76 & 1.00 & 0.76 \\
16260 & & 12 37 00.46 & $+$62 15 08.86 & 2.33 & 24.82 & 10.92 & 1.14 & 0.77 \\
10404 & & 02 17 35.93 & $-$05 13 05.20 & 2.31 & 22.92 & 10.04 & 1.55 & 0.51 \\
12138 & & 02 17 48.89 & $-$05 12 30.38 & 2.25 & 23.26 & 9.98 & 1.24 & 0.28 \\
14004 & & 02 17 30.72 & $-$05 11 56.22 & 2.39 & 23.25 & 10.46 & 1.54 & 0.57 \\
14042 & & 02 17 32.78 & $-$05 11 55.93 & 2.15 & 22.65 & 9.97 & 1.52 & 0.27 \\

\enddata 
\vspace{0.3cm}
\tablecomments{Observational measurements and derived physical properties for the SIGMA galaxy sample. The galaxy ID, RA (J2000) and DEC (J2000) are matched with the publicly available CANDELS catalogs in \citet{2013ApJS..206...10G, 2013ApJS..207...24G}. The H-band magnitude (m$_{H}$) and axis ratio (b/a) are adopted from the GALFIT measurements on the HST/WFC3 F160W CANDELS image in \citet{2012ApJS..203...24V}.  The stellar mass (M$_*$) and star-formation rate (SFR) are derived from the full galaxy SED, as described in \S3.2.} 
\label{tab:table1}
\end{deluxetable*}

\begin{deluxetable*}{ccccccccc} 
\tabletypesize{\footnotesize} 
\tablecolumns{9} 
\tablewidth{0pt} 
\tablecaption{SIGMA - Observational and Kinematic Properties}
\tablehead{ID  & \colhead{N$_{slits}$} &  $PA_{phot}$($^{\circ}$)&  $PA_{slit}$($^{\circ}$) & $\Delta\,PA$ & $r_v ($\arcsec$)$ & $V\,\times\,\sin i$ &$V_{rot}$ (km s$^{-1}$) & $\sigma_g$ (km s$^{-1}$)
}
\vspace{0.3cm}
\startdata 
5492  & 1 & 65.0    $\pm$ 0.7 & 34.5 & 30.5 &  0.2 & 10 $\pm$ 30 & 11 $\pm$ 35 & 45 $\pm$ 14 \\
14602 & 1 & $-$11.7 $\pm$ 0.7 & 19.0 & 30.7 &  0.2 & 25 $\pm$ 37 & 59 $\pm$ 87 & 85 $\pm$ 23 \\
16209 & 2 & 9.4     $\pm$ 1.3 & 19.0 & 9.6 &  0.2 & 65 $\pm$ 28 & 67 $\pm$ 29 & 86 $\pm$ 22 \\
16600 & 2 & 7.0     $\pm$ 0.2 & 10.0 & 3.0 &  0.2 & 171 $\pm$ 23 & 234 $\pm$ 31 & 35 $\pm$ 35 \\
16905 & 3 & $-$87.0 $\pm$ 0.3 & -56.0 & 31.0 &  0.4 & 131 $\pm$ 63 & 135 $\pm$ 65 & 80 $\pm$ 24 \\
17488 & 2 & 17.4    $\pm$ 1.8 & 19.0 & 1.6 &  0.2 & 5 $\pm$ 14 & 5 $\pm$ 15 & 45 $\pm$ 5 \\
17673 & 1 & $-$2.8  $\pm$ 0.6 & 8.0 & 10.8 &  0.2 & 44 $\pm$ 20 & 81 $\pm$ 37 & 35 $\pm$ 15 \\
18177 & 2 & 83.1    $\pm$ 0.4 & 56.0 & 27.1 &  0.4 & 161 $\pm$ 50 & 172 $\pm$ 53 & 90 $\pm$ 26 \\
19818 & 2 & $-$86.6 $\pm$ 1.4 & 56.0 & 37.4 &  0.3 & 24 $\pm$ 21 & 40 $\pm$ 35 & 80 $\pm$ 13 \\
20495 & 3 & 74.1    $\pm$ 0.3 & 56.0 & 18.1 &  0.4 & 120 $\pm$ 27 & 187 $\pm$ 42 & 65 $\pm$ 22 \\
20569 & 1 & 22.5    $\pm$ 1.8 & 8.0 & 14.5 &  0.2 & 31 $\pm$ 27 & 34 $\pm$ 30 & 36 $\pm$ 15 \\
20593 & 2 & 34.1    $\pm$ 0.9 & 19.0 & 15.1 &  0.4 & 130 $\pm$ 44 & 133 $\pm$ 45 & 49 $\pm$ 26 \\
20883 & 1 & 29.8    $\pm$ 2.1 & 8.0 & 21.8 &  0.2 & 51 $\pm$ 36 & 58 $\pm$ 41 & 89 $\pm$ 14 \\
21007 & 5 & $-$59.5 $\pm$ 0.4 & 117.0 & 3.5 &  0.2 & 136 $\pm$ 24 & 227 $\pm$ 40 & 39 $\pm$ 40 \\
21162 &  3 & $-$82.7 $\pm$ 0.7 & 19.0 & 78.3 & 0.2 &  52 $\pm$ 19 & --- & 80 $\pm$ 6 \\
21258 & 3 & $-$52.7 $\pm$ 1.4 & 117.0 & 10.3 &  0.4 & 101 $\pm$ 78 & 104 $\pm$ 80 & 55 $\pm$ 38 \\
21737 & 1 & $-$19.6 $\pm$ 0.6 & 8.0 & 27.6 &  0.2 & 157 $\pm$ 29 & 212 $\pm$ 39 & 30 $\pm$ 31 \\
21897 & 2 & $-$24.5 $\pm$ 0.7 & 120.5 & 35.0 &  0.2 & 136 $\pm$ 20 & 146 $\pm$ 21 & 55 $\pm$ 17 \\
22236 & 2 & $-$15.7 $\pm$ 1.5 & 19.0 & 34.7 &  0.2 & 110 $\pm$ 32 & 123 $\pm$ 35 & 50 $\pm$ 23 \\
22354 & 2 & 36.0    $\pm$ 0.7 & 19.0 & 17.0 &  0.2 & 56 $\pm$ 20 & 132 $\pm$ 47 & 30 $\pm$ 20 \\
22783 & 1 & 32.1    $\pm$ 1.5 & 10.0 & 22.1 &  0.2 & 5 $\pm$ 38 & 5 $\pm$ 41 & 40 $\pm$ 16 \\
23095 & 2 & $-$70.9 $\pm$ 0.5 & 120.5 & 11.4 &  0.2 & 200 $\pm$ 42 & 301 $\pm$ 63 & 34 $\pm$ 36 \\
23632 & 2 & 88.3    $\pm$ 1.2 & 78.5 & 9.8 &  0.2 & 34 $\pm$ 18 & 35 $\pm$ 18 & 39 $\pm$ 8 \\
25715 & 1 & $-$39.7 $\pm$ 0.9 & 120.5 & 19.8 &  0.2 & 75 $\pm$ 43 & 262 $\pm$ 150 & 109 $\pm$ 23 \\
25973 & 3 & 33.0    $\pm$ 0.8 & 19.0 & 14.0 &  0.2 & 79 $\pm$ 21 & 98 $\pm$ 26 & 40 $\pm$ 15 \\
26050 & 1 & $-$53.0 $\pm$ 0.6 & 117.0 & 10.0 &  0.3 & 222 $\pm$ 33 & 224 $\pm$ 33 & 50 $\pm$ 24 \\
26177 & 2 & $-$20.1 $\pm$ 0.2 & 120.5 & 39.4 &  0.2 & 74 $\pm$ 18 & 174 $\pm$ 42 & 45 $\pm$ 13 \\
26736 & 2 & $-$35.2 $\pm$ 0.5 & -10.0 & 25.2 &  0.2 & 242 $\pm$ 35 & 272 $\pm$ 39 & 75 $\pm$ 42 \\
1826  & 1 & $-$11.7 $\pm$ 1.2 & 32.0 & 43.7 &  0.3 & 30 $\pm$ 30 & 31 $\pm$ 31 & 49 $\pm$ 12 \\
2815  & 2 & $-$12.8 $\pm$ 3.4 & 32.0 & 44.8 &  0.3 & 29 $\pm$ 40 & 37 $\pm$ 52 & 64 $\pm$ 12 \\
4106  &  2 & $-$29.5 $\pm$ 0.4 & 32.0 & 61.5 & 0.3 &  82 $\pm$ 64 & --- & 129 $\pm$ 25 \\
4476  & 1 & 68.0    $\pm$ 2.4 & 106.9 & 38.9 &  0.3 & 90 $\pm$ 37 & 91 $\pm$ 37 & 60 $\pm$ 25 \\
4714  & 1 & $-$47.9 $\pm$ 1.2 & 106.9 & 25.2 &  0.3 & 68 $\pm$ 38 & 119 $\pm$ 66 & 84 $\pm$ 30 \\
4925  & 3 & $-$29.0 $\pm$ 1.2 & -21.5 & 7.5 &  0.2 & 318 $\pm$ 35 & 448 $\pm$ 49 & 25 $\pm$ 61 \\
4962  & 3 & 24.9    $\pm$ 3.9 & 32.0 & 7.1 &  0.2 & 124 $\pm$ 25 & 150 $\pm$ 30 & 55 $\pm$ 20 \\
9157  &  3 & 14.6    $\pm$ 0.8 & 106.9 & 87.7 & 0.3 &  64 $\pm$ 22 & --- & 85 $\pm$ 15 \\
9190  &  3 & $-$84.2 $\pm$ 1.6 & 32.0 & 63.8 & 0.2 &  21 $\pm$ 28 & --- & 36 $\pm$ 18 \\
10230 & 1 & $-$8.3  $\pm$ 7.7 & -53.0 & 44.7 &  0.3 & 50 $\pm$ 26 & 50 $\pm$ 26 & 45 $\pm$ 20 \\
11525 & 2 & 44.7    $\pm$ 2.7 & 32.0 & 12.7 &  0.2 & 256 $\pm$ 32 & 276 $\pm$ 34 & 10 $\pm$ 52 \\
13230 & 1 & 22.1    $\pm$ 18.5 & -21.5 & 43.6 &  0.3 & 10 $\pm$ 22 & 10 $\pm$ 23 & 75 $\pm$ 7 \\
13678 & 1 & $-$34.9 $\pm$ 0.2 & -53.0 & 18.1 &  0.3 & 60 $\pm$ 25 & 64 $\pm$ 26 & 61 $\pm$ 8 \\
14428 & 1 & $-$68.3 $\pm$ 5.6 & -53.0 & 15.3 &  0.2 & 82 $\pm$ 24 & 157 $\pm$ 46 & 71 $\pm$ 14 \\
15497 & 1 & 7.9     $\pm$ 21.9 & 32.0 & 24.1 &  0.2 & 49 $\pm$ 31 & 73 $\pm$ 46 & 54 $\pm$ 20 \\
16028 & 1 & $-$71.4 $\pm$ 5.8 & -53.0 & 18.4 &  0.2 & 203 $\pm$ 36 & 306 $\pm$ 54 & 110 $\pm$ 36 \\
16260 & 1 & $-$63.0 $\pm$ 10.3 & -53.0 & 10.0 &  0.2 & 200 $\pm$ 48 & 307 $\pm$ 73 & 85 $\pm$ 51 \\
10404 & 1 & 52.2    $\pm$ 1.8 & -297.0 & 10.8 &  0.2 & 100 $\pm$ 33 & 114 $\pm$ 37 & 74 $\pm$ 14 \\
12138 & 1 & 67.3    $\pm$ 1.9 & -297.0 & 4.3 &  0.1 & 46 $\pm$ 17 & 46 $\pm$ 17 & 55 $\pm$ 10 \\
14004 & 1 & 40.6    $\pm$ 1.3 & -297.0 & 22.4 &  0.2 & 248 $\pm$ 68 & 296 $\pm$ 81 & 15 $\pm$ 83 \\
14042 & 1 & 34.9    $\pm$ 0.9 & -297.0 & 28.1 &  0.3 & 50 $\pm$ 26 & 50 $\pm$ 26 & 70 $\pm$ 8 \\

\vspace{0.2cm}
\tablecomments{The observational parameters and best-fit kinematic properties for the SIGMA galaxy sample. N$_{slits}$ is the number of spectra in the full SIGMA sample. $\Delta\,PA$ is the difference in position angle between the H-band continuum ($PA_{phot}$) and the Keck/MOSFIRE slit selected for this paper (i.e., the slit with the maximum rotation signature; $PA_{slit}$). $r_v$ is the fixed turnover radius of the rotation curve model. $V\,\times\,\sin i$ is measured from the flat part of the best-fit rotation curve and $V_{rot}$ is the inclination corrected rotation velocity. $\sigma_g$ is the best-fit gas velocity dispersion and is assumed constant across the face of the galaxy.} 
\label{tab:table2}
\end{deluxetable*}

\newpage
\begin{figure*}
\begin{centering}
\includegraphics[angle=0,scale=.50]{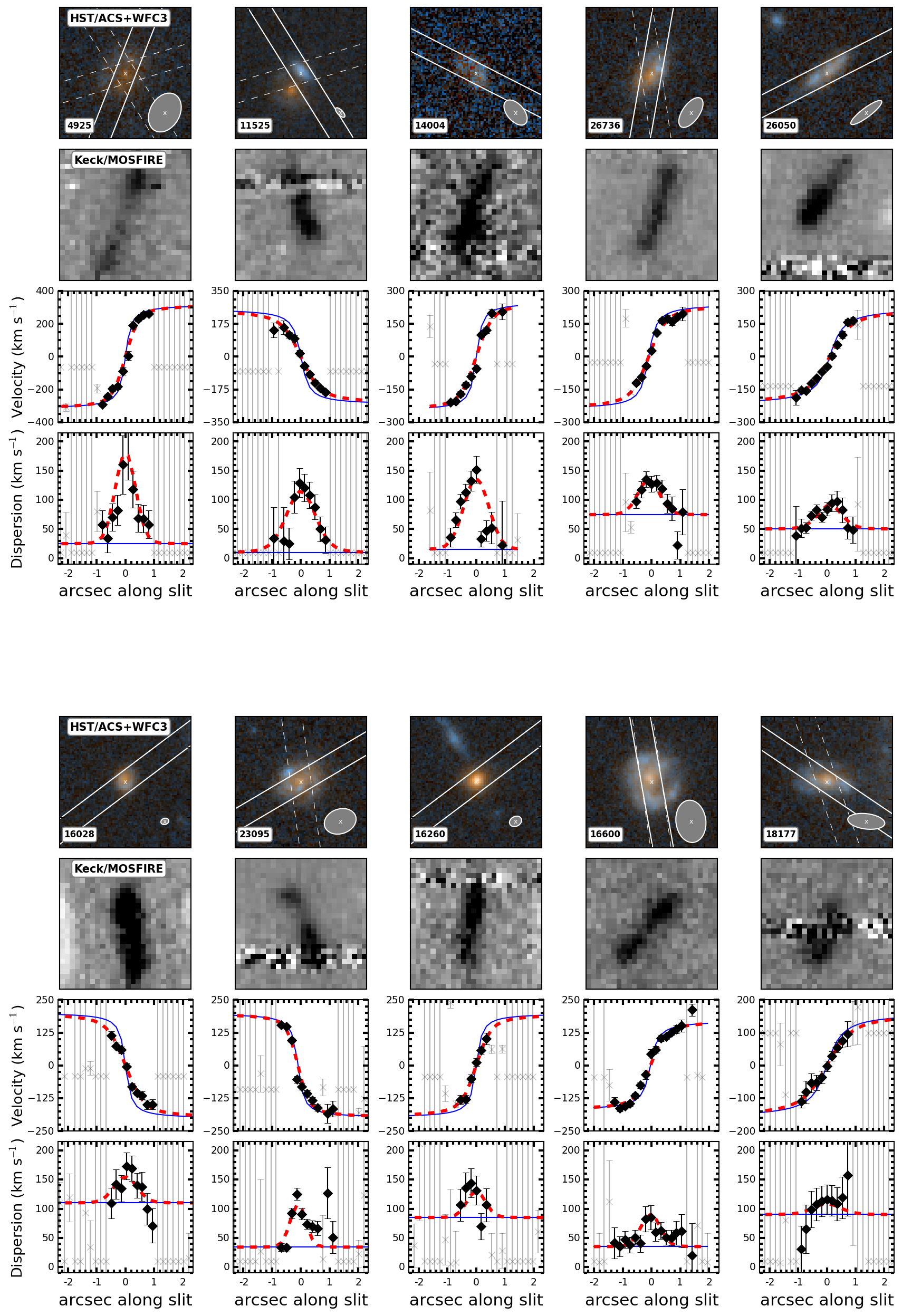}
\caption {Kinematic fits are presented for the full sample in this paper. Galaxies are ordered from largest to smallest rotation signature. In the top row, we show the CANDELS I+H-Band HST/ACS-WFC3 color image with the placements of the MOSFIRE slits and the {\tt GALFIT} best-fit half-light ellipse from the H-band image. The solid slit has been adopted for this paper. The Keck/MOSFIRE spectrum of the solid slit, centered around \OIII$\lambda$5007 or H$\alpha$, is shown in the second row. The typical seeing FWHM for these observations is 0.6$\arcsec$. The best-fit velocity and velocity dispersion models are shown as red solid lines and the intrinsic models are shown as blue dashed lines. The black filled diamonds represent the gaussian fits to the velocity and velocity dispersion in each row and the grey points are discarded fits. All of the rows are spatially aligned and each panel is 4.5$\arcsec$ on a side.}
\label{fig:full_data_fig1}
\end{centering}
\end{figure*}

\newpage
\begin{figure*}
\begin{centering}
\includegraphics[angle=0,scale=.50]{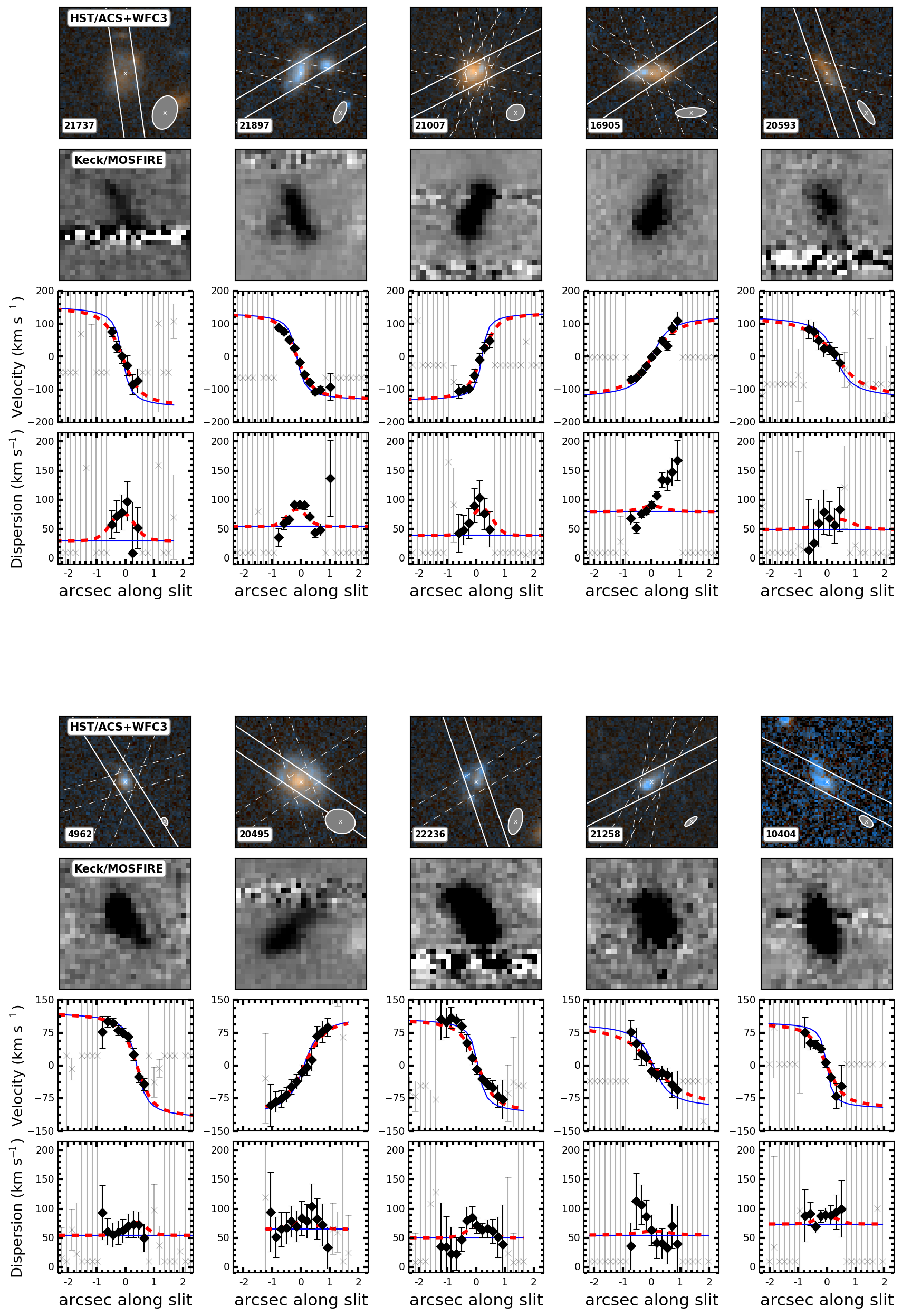}
\caption {Continuation of Figure \ref{fig:full_data_fig1}.}
\label{fig:full_data_fig2}
\end{centering}
\end{figure*}

\newpage
\begin{figure*}
\begin{centering}
\includegraphics[angle=0,scale=.50]{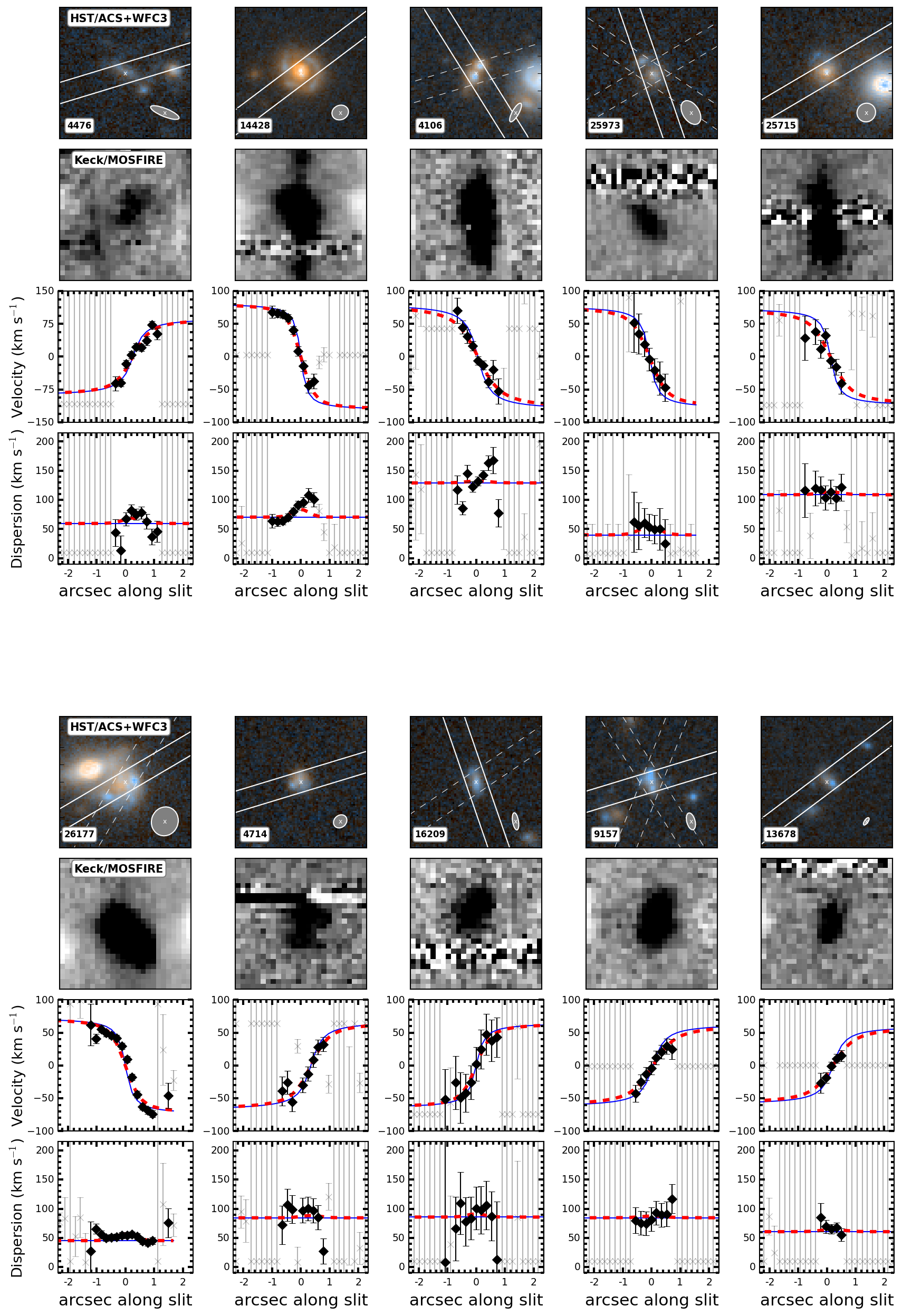}
\caption {Continuation of Figure \ref{fig:full_data_fig1}.}
\label{fig:full_data_fig3}
\end{centering}
\end{figure*}

\newpage
\begin{figure*}
\begin{centering}
\includegraphics[angle=0,scale=.50]{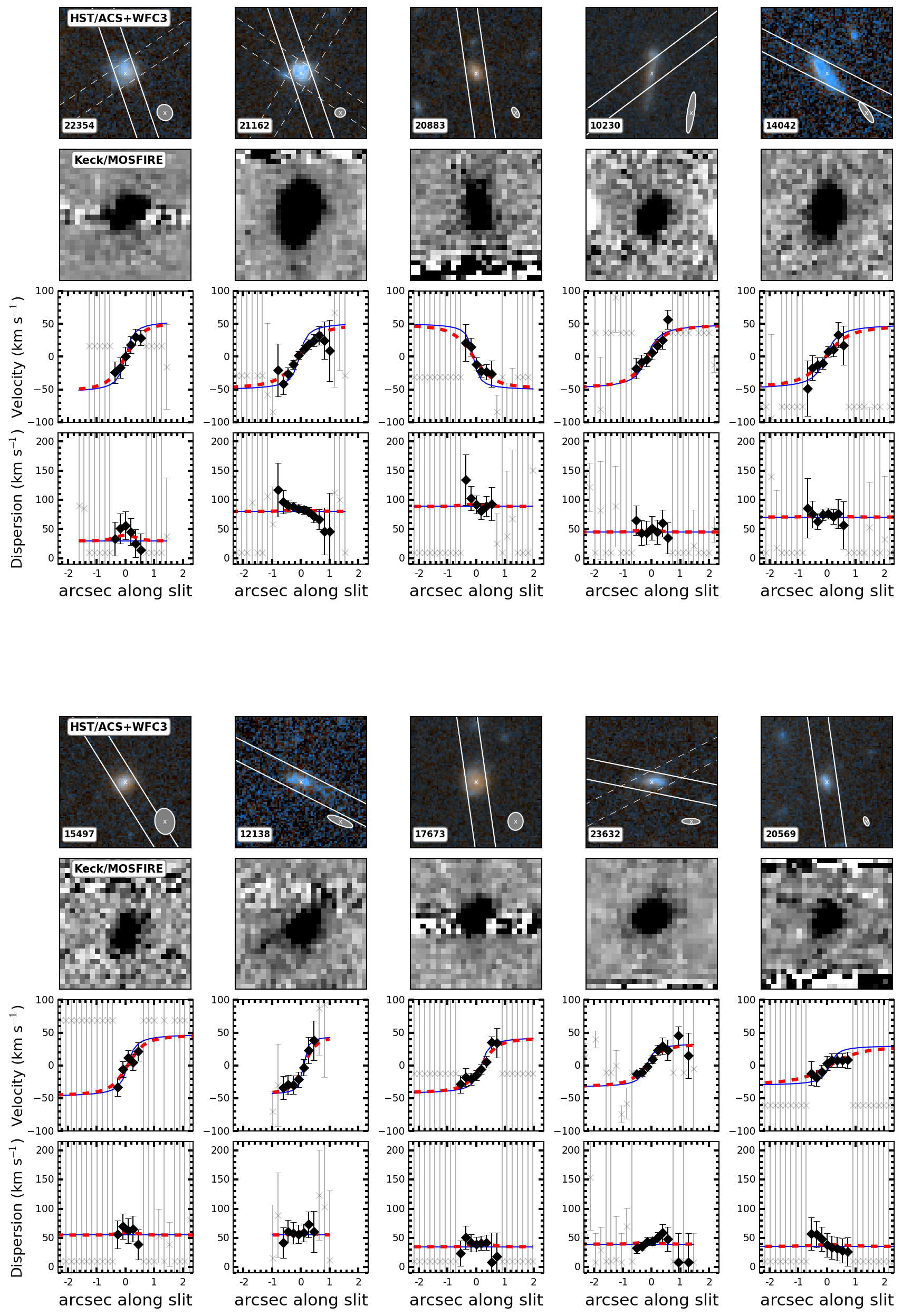}
\caption {Continuation of Figure \ref{fig:full_data_fig1}.}
\label{fig:full_data_fig4}
\end{centering}
\end{figure*}

\newpage
\begin{figure*}
\begin{centering}
\includegraphics[angle=0,scale=.50]{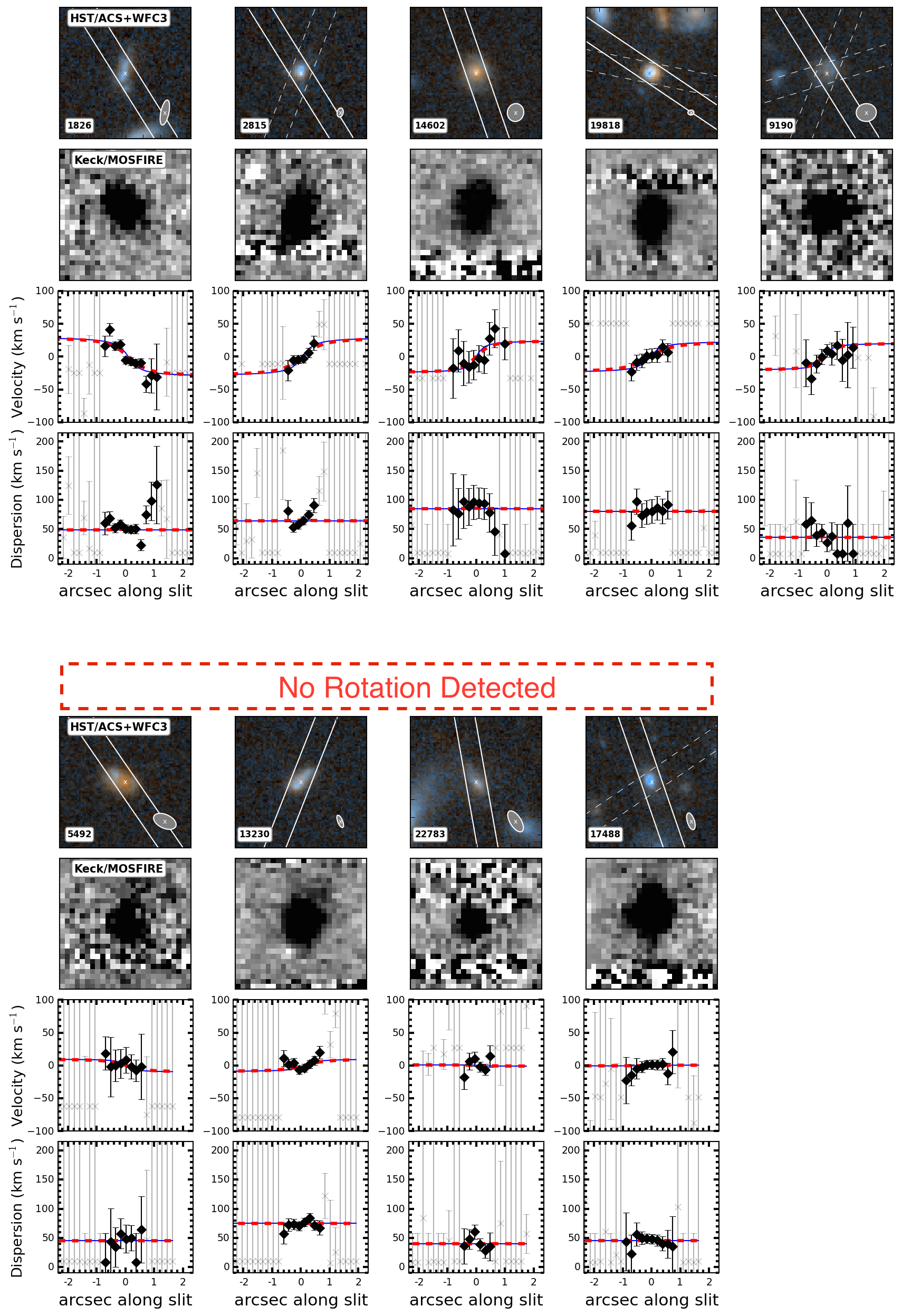}
\caption {Continuation of Figure \ref{fig:full_data_fig1}.}
\label{fig:full_data_fig5}
\end{centering}
\end{figure*}

\end{document}